\documentclass[aps,prd,12pt,nofootinbib,nobibnotes,eqsecnum,secnumarabic,superscriptaddress,tightenlines,notitlepage,preprintnumbers]{revtex4-1}
\usepackage[lofdepth,lotdepth,caption=false]{subfig}

\usepackage{amsmath, amssymb, graphicx}
\usepackage{palatino}
\usepackage{xspace}
\usepackage{braket}
\usepackage{color}

\DeclareMathOperator{\Tr}{\mathrm{Tr}}

\newcommand\bigO{O} 
\newcommand\vev[1]{\left\langle #1 \right\rangle}
\newcommand\dg{\dagger}
\newcommand\no[1]{:\hspace{-2pt} #1 \hspace{-1pt}:\,}

\newcommand{\drm}{\mathrm{d}}
\newcommand{\der}[2]{\frac{\drm #1}{\drm #2}}
\newcommand{\pd}{\partial}
\newcommand{\re}{\mathbb{R}}
\newcommand{\co}{\mathbb{C}}
\newcommand{\ints}{\mathbb{Z}}

\newcommand{\beq}{\begin{equation}}
\newcommand{\eeq}{\end{equation}}

\newcommand{\abs}[1]{\left\lvert#1\right\rvert}

\makeatletter
\newenvironment{calc}{\allowdisplaybreaks\start@align\@ne\st@rredtrue\m@ne}
{\addtocounter{equation}{1}\tag{\theequation}\endalign}
\makeatother

\begin{document}
\preprint{NSF-KITP-11-179}
\preprint{IMSc/2011/8/7}
\title{Evolution of entanglement entropy in the D1--D5 brane system}
\author{Curtis T. Asplund}
\email{casplund@physics.ucsb.edu}
\affiliation{Dept. of Physics, University of California, Santa Barbara, California 93106, USA}

\author{Steven G. Avery}
\email{avery@mps.ohio-state.edu}
\affiliation{The Institute of Mathematical Sciences, CIT Campus, Taramani, Chennai, India 600113}
\affiliation{Kavli Institute for Theoretical Physics, University of California, Santa Barbara, California 93106, USA}
\begin{abstract}
  We calculate the evolution of the geometric entanglement entropy following a local quench in the D1D5 conformal field theory, a two-dimensional theory that describes a particular bound state of D1 and D5 branes.  The quench corresponds to a localized insertion of the exactly marginal operator that deforms the field theory off of the orbifold (free) point in its moduli space. This deformation ultimately leads to thermalization of the system.  We find an exact analytic expression for the entanglement entropy of any spatial interval as a function of time after the quench and analyze its properties. This process is holographically dual to one stage in the formation of a stringy black hole.
\end{abstract}
\maketitle

\section{Introduction}

Consider an initial, smooth configuration of matter that collapses
into a black hole. There are longstanding questions about how the
information in the initial configuration, such as the entanglement
between various subsystems, becomes encoded in the resulting black
hole. As a quantum theory of gravity, string theory addresses many of
these questions. While the AdS/CFT correspondence leads immediately to
the proposal that certain black holes are dual to thermal mixed states
of a dual conformal field theory (CFT) \cite{maldacena_large_1998},
this says little about the formation process. To go further toward
answering such dynamical questions, one needs to study the unitary
evolution of a CFT with a gravitational dual, undergoing
thermalization. In this paper we begin such an investigation in the
D1D5 CFT, which describes a bound state of D1 and D5 branes, and is
well-known as a useful system for studying black holes in string
theory.
	
We study this process using the entanglement entropy, defined below,
which measures entanglement between subsystems in a quantum system.
As opposed to the many studies of black hole entropy as entanglement
entropy of different parts of the bulk spacetime or between different
boundary CFTs (\cite{solodukhin_entanglement_2011} reviews many of
these), we are considering the evolution of the entanglement entropy
of subsystems of a single CFT, in order to study its thermalization.
The connection between quantum entanglement and thermodynamics has a
long history. See~\cite{lloyd_quantum_2006} for a short review,
\cite{deutsch_quantum_1991, srednicki_chaos_1994} for relevant early
investigations and \cite{popescu_entanglement_2006,
  gell-mann_quasiclassical_2007, rigol_relaxation_2007-1,
  rigol_thermalization_2008, rigol_alternatives_2011} for recent
general results.
  
As a brief review, begin with a quantum system with Hilbert space
$\mathcal{H}$ and Hamiltonian $H$.  Then factorize, or coarse-grain,
the Hilbert space of the full system as $\mathcal{H} = \mathcal{H}_{A}
\otimes \mathcal{H}_{E}$, where the first factor contains all states
describing degrees of freedom in a subsystem $A$ and the second factor
all states for the exterior of $A$---the ``environment'' $E$. We call a
quantum system, in a state specified by a density matrix $\rho$,
thermalized if for general small subsystems $A$ the reduced density
matrices $\rho_{A}$, obtained by tracing over $\mathcal{H}_{E}$, are
approximately (Gibbs/canonical ensemble) thermal mixed states, e.g.,
\beq
\label{eq:thermst}
\rho_{A} \approx \frac{e^{-\beta H_{A}}}{Z_{A}} \eeq for some $\beta$
(the inverse temperature), where $H_{A} = \Tr_{E} H$ and $Z_{A} =
\Tr_{A} e^{-\beta H_{A}}$. This is an admittedly imprecise definition 
of ``thermalized,'' but is sufficient for our purposes. The expression in 
\eqref{eq:thermst} is appropriate for those
cases where the Hamiltonian is the only conserved quantity. For
systems with more symmetries, including integrable systems with
infinitely many conserved quantities, one can still define generalized
Gibbs ensembles and appropriate thermal reduced density matrices,
characterized by generalized chemical potentials in addition to
$\beta$ \cite{popescu_entanglement_2006, rigol_relaxation_2007-1}. 
This is relevant for CFTs like the one we
consider, which undergo a quench but which subsequently evolve as free
theories and are characterized by a set of momentum-dependent
temperatures \cite{calabrese_quantum_2007, rigol_alternatives_2011}. 

We are concerned here with the case that the full system $AE$ is in a
pure state, i.e., a closed quantum system.  Then the fact that
$\rho_{A}$ is mixed comes entirely from the entanglement of $A$ with
$E$. This entanglement can be measured in a variety of ways, but in
this paper we consider the R\'enyi and von Neumann entropies, which
are given by \eqref{eq:renyi} below. We choose these quantities for a
number of reasons, including nice analytic properties and
calculability, to be discussed in the paper.  Here we just emphasize
that they let one track the thermalization of various subsystems as
well as deviations from strictly thermal behavior (see
\cite{susskind_introduction_2004} \S 8.2 for an introductory
discussion of this topic). The entanglement entropies, as a function
of subsystem, time, and other parameters, can also yield much other
information about the system. This is explored in voluminous recent
work in condensed matter physics (see \cite{_entanglement_2009} for
several recent reviews).

One can use these quantities, in principle, to investigate the
recently conjectured thermalization time for black holes that
saturates a causality bound \cite{sekino_fast_2008}, although we don't
get that far in this paper.  As we discuss below, it is technically
difficult to quantitatively compare the nonequilibrium dynamics we
study here to the system in equilibrium and the full thermalization
process. However, we can still learn a lot from the results we
present.

Motivated by the rapid thermalization observed in heavy ion collisions
as well as the theoretical questions already mentioned, there are many
investigations that use AdS/CFT to study strongly coupled CFTs far
from equilibrium or undergoing thermalization, e.g., \cite{Das:2010yw,
  danielsson_black_2000, giddings_gravitational_2002,
  chesler_horizon_2009, bhattacharyya_weak_2009, asplund_small_2009,
  hubeny_holographic_2010}. Some of these
\cite{balasubramanian_thermalization_2011,
  balasubramanian_holographic_2011, albash_evolution_2011,
  abajo-arrastia_holographic_2010} also use entanglement entropy via
the holographic entanglement entropy proposal
\cite{ryu_holographic_2006, hubeny_covariant_2007} (see also
\cite{headrick_entanglement_2010} for a recent study of the
holographic R\'enyi entropies). These latter investigations use the
dual classical geometry and so cannot address some of the most
puzzling questions about information in black holes, which involve
quantum mechanics of the bulk theory in an essential way.
	 
We study the D1D5 CFT at weak coupling, which has long been used in
string theory to study black holes \cite{strominger_microscopic_1996,
  callan_d-brane_1996}. Early studies focused on the extremal,
zero-temperature configurations, whereas we consider exciting to a
state that is far from extremal. Recent work has studied Hawking
radiation in this system in detail \cite{chowdhury_radiation_2008,
  chowdhury_pair_2008, chowdhury_non-extremal_2009,
  avery_emission_2009, avery_emission_2010} and deformations of the
CFT away from the orbifold (free) point in the moduli space
\cite{avery_deforming_2010, avery_excitations_2010,
  avery_intertwining_2011} (see also \cite{gava_proving_2002,
  Pakman:2009mi} for leading order calculations away from the orbifold
point). Because we work at weak coupling, i.e.  near the orbifold
point, we are not in the regime where supergravity is a good
approximation. Nonetheless the above work indicates that this regime
contains much information about black holes. Additional support for
this includes several precise matchings between gravitational
calculations and calculations from the free theory data
\cite{das_comparing_1996, birmingham_conformal_2002,
  birmingham_relaxation_2003}.  More evidence for the surprising
efficacy of the orbifold CFT in describing black holes comes from the
very recent paper~\cite{bena_moulting_2011}.  There are also general
arguments for thermalization of CFTs with gravity duals in the large
$N$ limit at any finite value of the coupling
\cite{festuccia_arrow_2007} corroborated by exact calculations in
simplified models \cite{iizuka_matrix_2008, iizuka_matrix_2010}. The
resulting weakly coupled thermal state is dual to a ``stringy black
hole,'' in that the string length is large compared to the size of the
black hole (see \cite{festuccia_arrow_2007} for further discussion of
stringy black holes).  The above work indicates that such a state
should also tell us about traditional black holes.

In particular, the results illustrated in \S\ref{sec:sample} have a
natural explanation in terms of free excitations traveling at the
speed of light around the $S^1$. This picture is similar to the CFT
description in~\cite{Lunin:2001dt, Giusto:2004ip} of near-extremal
supergravity excitations, which, when the decoupling limit is relaxed,
 can periodically escape the AdS throat to the asymptotic flat space.
The period and rate of emission were reproduced from the same kind of
CFT dynamics we observe. Thus our results correctly
capture some qualitative aspects of the supergravity description. On
the other hand, we expect that large energy (far from extremal)
supergravity excitations can back-react to form black holes. In the
CFT, this corresponds to thermalization. Since the entanglement
entropy that we find does not persist, but rather has short-time
periodic dynamics, we conclude that, as expected, the orbifold
CFT does not capture this important process. We hope to
address this issue more quantitatively in future works.
	 
A closely related precursor to our work is
\cite{takayanagi_measuring_2010}, which also calculates the evolution
of entanglement entropy in a weakly coupled CFT. The authors proposed this as
a way to study quantum black hole formation and emphasized that the
entanglement entropy can be thought of as a coarse-grained
thermodynamic entropy. However, their CFT (a single fermion) has no
clear dual black hole interpretation, although it does illustrate
some general features of the kind of problem we are considering.

The general process of
thermalization in weakly coupled theories is well studied, but the
D1D5 system exhibits some novel features. In particular, we consider dynamics
 arising from a localized insertion of
 a particular marginal deformation of the
orbifold CFT 
that acts as a \emph{local
  quench}, to be described below. The calculation of the entropy
produced by this quench is the main result of this paper. This is
the basic process by which entropy is generated. [Note added in proof: 
After this article was written we learned of \cite{2011JSMTE..08..019S}, which contains calculations of similar quantities and appears consistent with our results.]

As emphasized in \cite{festuccia_arrow_2007}, the familiar
semiclassical dynamics of black holes, including the puzzling
apparent loss of information, should appear in the limit $N
\rightarrow \infty$ of the dual holographic theory.  Here that would
correspond to the limit of infinitely many D1 and D5 branes.  We do
not consider that limit here, rather we consider a finite number of
branes unitarily evolving toward a thermalized pure state as described
above.  In this paper we just analyze the basic process in that
evolution.
	 	 
In \S\ref{sec:D1D5} we review the D1D5 CFT and the marginal
deformations that we study. In \S\ref{sec:setup} we set up the
calculation of the entanglement entropies in the CFT, including a
review of the replica trick. Next, in \S\ref{sec:4pt} we compute the
four-point function that we need to calculate the entropies, which we
do in \S\ref{sec:entropy}. We illustrate some of their properties in
\S\ref{sec:sample}. We conclude with a discussion of our results and
future directions.

\section{D1D5 review}	
\label{sec:D1D5}
	
The D1D5 system is realized in IIB string theory compactified
on\footnote{One may also consider K3 instead of $T^4$.}  $T^4\times
S^1$ with the bound state of $N_1$ D1-branes wrapping the $S^1$ and
$N_5$ D5-branes wrapping $T^4\times S^1$. We take the $S^1$ to be
large compared to the $T^4$.  The near-horizon limit of the geometry
is $AdS_3\times S^3\times T^4$, which is dual to a two-dimensional CFT
living on the boundary of $AdS_3$.

The two-dimensional D1D5 CFT has $\mathcal{N} = (4,4)$ supersymmetry
with $\mathrm{SU}(2)_L\times \mathrm{SU}(2)_R$ R-symmetry corresponding to the isometry
of the $S^3$. The two-dimensional base space of the CFT is given by
the cylindrical boundary of $AdS_3$ parametrized by time and the
$S^1$. In addition, we can organize the field content using the
$\mathrm{SO}(4)_I\simeq \mathrm{SU}(2)_1\times \mathrm{SU}(2)_2$ symmetry broken by the
compactification on $T^4$. One can also fix the total central charge
$c=6N_1N_5$ from the algebra of diffeomorphisms that preserve the
asymptotic $AdS_3$. The CFT has a twenty-dimensional moduli space that
corresponds to the near-horizon twenty-dimensional moduli space of the
IIB supergravity compactification.

There is a point in moduli space called the ``orbifold point,''
analogous to free SYM in $\text{AdS}_5/\text{CFT}_4$, where the D1D5 CFT is a sigma
model with orbifolded target space, $(T^4)^{N_1N_5}/S_{N_1N_5}$. 
Just
as the dual of free SYM does not have a good geometric description,
the orbifold CFT is far from points in moduli space that are well
described by supergravity.  We wish to study the effect of certain
$(4,4)$ exactly marginal deformations that move the orbifold CFT
toward points in moduli space that should have geometric descriptions
and, in particular, should include black hole physics. Even
though we work far from the supergravity regime, as discussed in the introduction, we can still capture some black hole physics.


We can think of the orbifold model as $N_1N_5$ copies of a $(4,4)$
$c=6$ CFT. Each copy has four real bosons that are vectors of
$\mathrm{SO}(4)_I$, $X^i$, and their fermionic superpartners. See,
e.g., \cite{avery_using_2010} for details. For computational purposes, we map
the real cylinder coordinates, $t\in\re$ and $y\in[0, 2\pi R)$, to
dimensionless complex coordinates on the cylinder
\begin{equation}
\label{eq:cyl}
w = \tau + i\sigma \qquad \frac{t}{R}\mapsto -i\tau\qquad \frac{y}{R} = \theta.
\end{equation}
Note that we have also incorporated a Wick rotation in this step. We
prefer to perform most of the computation in the complex plane by
further mapping to coordinates
\begin{equation}\label{eq:exp-map}
z = e^w\qquad \bar{z} = e^{\bar{w}}.
\end{equation}

In addition to the local bosonic and fermionic excitations of each
copy, the orbifold theory also has twisted sectors: states which come
back to themselves only up to an element of the orbifold group
$S_{N_1N_5}$ upon circling the $S^1$. The twist operators
$\sigma_n(z)$ are labeled by $n$-cycles and change the twist sector of
the theory. More concretely, consider operators $O^{(i)}(z)$ in the
$i$th copy. In the presence of $\sigma_{(12\dots n)}(z_0)$, the
operators have boundary conditions
\begin{equation}
O^{(i)}(z_0 + ze^{2\pi i}) = 
\begin{cases}
O^{(i+1)}(z_0+z) & i= 1,\dots, n-1\\
O^{(1)}(z_0 + z) & i=n\\
O^{(i)}(z_0 + z) & i= n+1, \dots, N_1N_5
\end{cases}.
\end{equation}
Let us emphasize that the twist operators considered here are physical
components of the orbifold CFT, and should not be confused with twist
operators introduced as part of the replica trick.

Following~\cite{avery_deforming_2010, avery_excitations_2010,
  avery_intertwining_2011}, we focus on four of the marginal
deformations that involve twist operators. These operators are
believed to be responsible for thermalization in the D1D5 CFT. The
$(4,4)$ supersymmetric deformations are singlets under $\mathrm{SU}(2)_L\times
\mathrm{SU}(2)_R$.  To obtain such a singlet we apply modes of the supercharges
$G^\mp_{\dot A}$ to $\sigma_2^\pm$, where we use plus/minus indices to
label elements of $\mathrm{SU}(2)_L$ doublets and dotted capital Latin indices
for doublets of $\mathrm{SU}(2)_2$.  In~\cite{avery_deforming_2010} it was
shown that we can write the deformation operator(s) as
\begin{equation}\label{eq:def-op}
\widehat O_{\dot A\dot B}(w_0)= 2
  \Big[\int_{w_0} \frac{\drm w}{2\pi i} G^-_{\dot A} (w)\Big]
  \Big[\int_{\bar w_0} \frac{\drm \bar{w}}{2\pi i} \bar G^-_{\dot B} (\bar w)\Big]
   \sigma_2^{++}(w_0),
\end{equation}
where the factor of $2$ normalizes the operator. The operator
$\sigma^+_2$ is normalized to have unit OPE with its conjugate
\begin{equation}
\sigma_{2,+}(z')\sigma_2^{+}(z)\sim \frac{1}{z'-z}.
\end{equation}
This implies that acting on the Ramond vacuum~\cite{avery_deforming_2010}
\begin{equation}
\sigma_2^+(z)|0_R^-\rangle^{(1)} |0_R^-\rangle^{(2)}=|0_R^-\rangle+\bigO(z).
\label{qwthree}
\end{equation}
Here $|0_R^-\rangle$ is the spin down Ramond vacuum of the CFT on the
doubly wound circle produced after the twist. The normalization
\eqref{qwthree} has given us the coefficient unity for the first term
on the RHS and the $\bigO(z)$ represent excited states of the
CFT on the doubly wound circle.

\section{Setup}
\label{sec:setup}

Let us now outline the precise calculation we perform. Since we are
interested in the dynamics of thermalization or scrambling in the CFT,
we quench the system and then look at the entanglement entropy of
spatial subsystems as a function of time. The entanglement entropy of
subsystems, as discussed, is a very natural quantity to examine when
discussing thermalization.  Happily, there is already some
considerable technology for computing the entanglement entropy after
both global and local quenches in two-dimensional CFTs \cite{calabrese_evolution_2005, calabrese_entanglement_2007}.

The specific quench we consider is a local insertion of the
deformation operator introduced above. Since this operator is believed
to be responsible for thermalization, it seems natural to consider the
dynamics after its application. Moreover, these results should tie in
strongly with previous investigations~\cite{avery_deforming_2010,
  avery_excitations_2010,avery_intertwining_2011}, which showed that
the deformation operator, in essence, effects a Bogolyubov
transformation. For instance, the deformation operator, when acting on
the vacuum, produces a squeezed state of the
form~\cite{avery_deforming_2010}
\begin{equation}
\label{eq:sqstate}
\sigma_2^+(z_0)\ket{0_R^-}^{(1)}\ket{0_R^{-}}^{(2)} =
\exp\left[
  -\frac{1}{2}\sum_{m, n}\gamma^B_{mn}\alpha_{A\dot{A},-m}\alpha^{A\dot{A}}_{-n}
+\sum_{m,n}\gamma^F_{mn}\psi^{+A}_{-m}{\psi^-}_{A,-n}\right]\ket{0^-_R}.
\end{equation}
In this equation, we only show the left (holomorphic) sector and
consider just the twist part of the deformation. On the left-hand side
we have the two-twist operator acting on the untwisted Ramond vacua,
which produces many pairs of bosonic and fermionic excitations on the
two-twisted Ramond vacuum. Note that the Ramond vacua have an
$\mathrm{SU}(2)_L\times \mathrm{SU}(2)_R$ spin structure.  The coefficients $\gamma^B$
and $\gamma^F$ are functions of $z_0$ given explicitly
in~\cite{avery_deforming_2010}. The calculation we propose, then,
computes the time dependence of the entanglement entropy of this
squeezed state; although, we do not use the above form explicitly.

The physical setup is as follows. We apply the deformation
operator~\eqref{eq:def-op} to the Ramond vacuum at time $t=0$ and
$y=0$. We then look at the time dependence of the entanglement entropy
of an arbitrary spatial interval. Since we are mostly interested in
how the entanglement entropy changes due to the quench, we subtract
off the entanglement entropy of the vacuum. We act in the Ramond
sector since that is the sector relevant for black holes.  We will
sketch the gravitational picture this corresponds to in the final
section.

\subsection{Review of the replica trick for computing entanglement
  entropy}
\label{sec:replica}

In the remainder of this section, we set up the calculation of the
entanglement entropy after the quench. Consider a system $S$ with some
subsystem $A$, and its complement $B$.  Recall that the (von Neumann)
entanglement entropy of $A$ in $S$ is defined as the von Neumann
entropy of the reduced density matrix,
\begin{equation}
\label{dm}
S(A) = -\Tr_A\hat{\rho}_A\log\hat{\rho}_A\qquad
\hat{\rho}_A = \Tr_{B} \hat{\rho}_S.
\end{equation}
The density matrix $\hat{\rho}_S$ is the density matrix for the full
system $S$. If, as is true throughout this paper, the total system $S$
is in a pure state $\ket{\psi}$, then $\hat{\rho}_S =
\ket{\psi}\bra{\psi}$. For our calculation, the subsystem $A$
corresponds to degrees of freedom living on some interval of $S^1$.
This definition has a number of nice properties that make it the
natural measure of entanglement 
including positive definiteness, strong subadditivity, and $S(A)= S(B)$ for a pure state. In
fact, this is essentially the unique measure of
entanglement satisfying the above properties \cite{ochs_new_1975}.

Computing the von Neumann entanglement entropy is computationally
difficult because of the log, so instead we follow
\cite{callan_geometric_1994, holzhey_geometric_1994,
  calabrese_entanglement_2004} and use the replica trick: we first
compute the R\'{e}nyi entropy of order $n$ and then analytically
continue to the von Neumann entropy. Recall that the R\'{e}nyi entropy
of order $n$ is defined as
\begin{equation}
\label{eq:renyi}
  S_n(A) = \frac{1}{1-n}\log\left(\Tr_{A} \hat{\rho}_A^n\right)\ \ \ \mbox{and}\ \ \ S_\text{vN}(A) = \lim_{n\to 1} S_n(A).
\end{equation}
The R\'{e}nyi entropies are an interesting measure of entanglement
even before taking the limit to the von Neumann entanglement entropy.
In particular, they serve as a lower bound on $S_{\mathrm{vN}}$
and vanish on an unentangled state.

Before showing how to compute the R\'{e}nyi entropy, we first review
how to write the density matrix $\hat{\rho}_S$ as a path integral.
From there we can easily compute $\Tr \rho_A^n$ as a path integral
with twisted boundary conditions. Let us work in some basis with
states that we will write as $\ket{\varphi}$; it is perhaps most
natural to think of these as shape states (field eigenstates), but any
basis works. Then, the $\varphi_1$--$\varphi_2$ element of the density
matrix at time $T$ can be written as
\begin{calc}
	\label{eq:dm}
  \bra{\varphi_2}\hat{\rho}(T)\ket{\varphi_1} 
   &= \braket{\varphi_2|\psi(T)}\braket{\psi(T)|\varphi_1}\\
   &= \bra{\varphi_2}e^{-i\hat{H}T}\ket{\psi_0}\bra{\psi_0}e^{i\hat{H}T}\ket{\varphi_1}\\
   &= \bra{\psi_0}e^{i\hat{H}T}\ket{\varphi_1}\bra{\varphi_2}e^{-i\hat{H}T}\ket{\psi_0},
\end{calc}
where we have suggestively switched the order of the two amplitudes
for reasons that should become clear. The state $\ket{\psi_0}$ is the
state at $t=0$, which for us is the state immediately after the
quench.  We have the product of two amplitudes, each of which can be
written as a separate path integral; however, it is more fruitful to
think of this as one path integral with discontinuous intermediate
boundary conditions. More specifically,
\begin{calc}
  \bra{\varphi_2}\hat{\rho}(T)\ket{\varphi_1} 
  &= \int D\phi(t)\bigg|_{\phi(-T) = \psi_0}^{\phi(0) = \varphi_2} e^{i\int_{-T}^0\drm t\, L(\phi(t))}
     \int D\phi(t)\bigg|_{\phi(0) = \varphi_1}^{\phi(-T) = \psi_0} e^{i\int_{0}^{-T}\drm t\, L(\phi(t))}\\
  &= \int D\phi(t)\bigg|_{\text{BCs}} e^{i \int_{C}\drm t\, L(\phi(t))},
\end{calc}
where ``BCs'' in the last line indicates the boundary conditions from
the previous line, and the contour $C$ starts at $t=-T$ goes to $t=0$
and then backwards to $t=-T$. We see, then, that we can think of the
density matrix as a path integral which accepts two boundary
conditions at $t=0$. We have translated the $\psi_0$ boundary
condition down to $t=-T$ to match with previous calculations~\cite{calabrese_evolution_2005, calabrese_entanglement_2007}.

\begin{figure}[ht]
\begin{center}
\includegraphics[width=6cm]{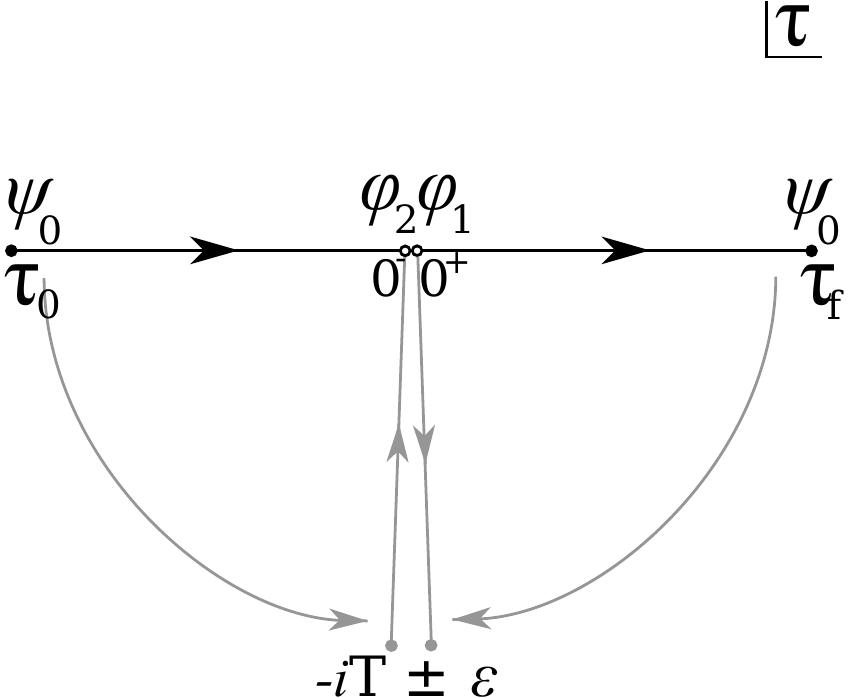}
\end{center}
\caption{The contour along the real axis of the complex $\tau$-plane
  for the Euclidean path integral. The analytic continuation back to
  Lorentzian time is shown in light gray, which shows how the
  $\epsilon$ regularization arises.}
 \label{contour}
\end{figure}

While this is formally correct, there are a couple of subtleties to
address.  First of all, we should clarify what we mean by the above
path integral, since as written we need double-valued fields. It is
more precise to parametrize $C$ as
\begin{equation}
t = \begin{cases} s & s\in[-T, 0]\\
 -s & s\in(0,T]
\end{cases}
\qquad s\in[-T,T],
\end{equation}
in which case the action in the path integral becomes
\begin{equation}\label{eq:contour-L}
\int_{C}\drm t\, L(\phi(t)) = \int_{-T}^0\drm s\,  L(\phi(s)) - \int_0^{T}\drm s\, L^T(\phi(s)),
\end{equation}
and $\phi$ is single-valued on $s$. We put the superscipt $T$ on $L$
in the second term to indicate that it is the time-reversed
Lagrangian. The second issue we need to address is Wick-rotating the
path integral. We usually Wick-rotate the path integral to imaginary
time to make the oscillatory term $iS$ into a convergent $-S_E$;
however, we now have a minus sign between the two terms
in~\eqref{eq:contour-L}, which means that we should Wick-rotate the
two terms oppositely, see Fig. \ref{contour}. When we Wick-rotate the second part in the
opposite direction, we get rid of the minus sign and the
time-reversal: we get a smoothly defined Euclidean path integral
\begin{equation}\label{eq:euc-path-integral}\begin{gathered}
Z(\tau_0, \tau_f; \psi_0; \varphi_1, \varphi_2) = 
    \int D\phi(\tau)\bigg|_{\text{BCs}}\exp\left({-\int_{\tau_0}^{\tau_f}\drm \tau\, L_E(\phi(\tau))}\right)\\
\text{BCs:}\quad \phi(\tau_0)=\phi(\tau_f) = \psi_0,\qquad 
    \phi(0^-)= \varphi_2, \qquad \phi(0^+) = \varphi_1.
\end{gathered}\end{equation}
We can compute this path integral for $\tau_0 < 0 < \tau_f$ with real
$\tau_0$ and $\tau_f$, and finally analytically continue to the
desired matrix element via
\begin{equation}
\bra{\varphi_2}\hat{\rho}\ket{\varphi_1} 
    = \lim_{\epsilon\to 0^+} N_\epsilon\,Z(-iT - \epsilon, -iT+\epsilon; \psi_0; \varphi_1, \varphi_2),
\end{equation}
where we put in $\epsilon$ to ``remember'' which direction we
Wick-rotated the two terms. The factor of $N_\epsilon$ is a
normalization constant that ensures $\Tr \hat{\rho} = 1$,
\begin{equation}
\frac{1}{N_\epsilon} = \bra{\psi_0}e^{-(2\epsilon)\hat{H}}\ket{\psi_0}
  = \int D\phi\,Z(-iT - \epsilon, -iT+\epsilon; \psi_0; \varphi, \varphi).
\end{equation}
The limit as $\epsilon\to 0$ is both delicate and crucial to getting
the right physics, since $\phi(\tau)$ has a branch cut along the
negative imaginary axis. Later, it should become clear that $\epsilon$
plays the role of a UV cutoff.

Before continuing, let us remark that the above should be reminiscent
of the Schwinger--Keldysh, or closed time path, formalism with temperature $T=
1/(2\epsilon)$ (see, e.g., \cite{calzetta_nonequilibrium_2008} for a review of this formalism). Indeed, if one integrates over $\psi_0$, then it is
exactly the Schwinger--Keldysh formalism, with some insertions at
$t=0$. Also note that if one identifies $\varphi_1 =\varphi_2 =
\varphi$ and integrates over $\varphi$, then one computes
$\Tr\hat{\rho}$, which is unity for a pure state.

\begin{figure}[ht]
\begin{center}
\includegraphics[width=8cm]{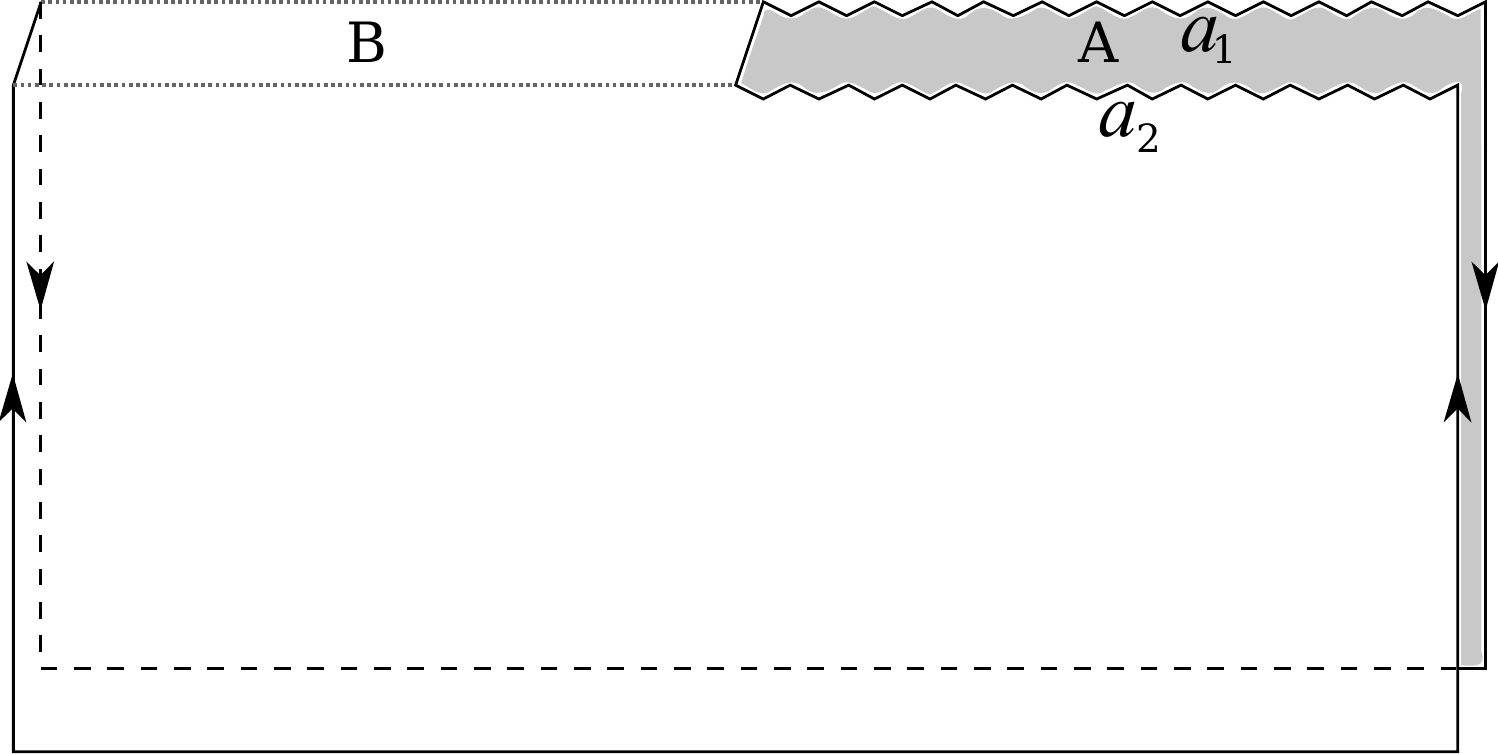}
\end{center}
\caption{The reduced density matrix as a path integral. Note that the flat
  piece on top is there for illustrative purposes only. The boundary
  conditions on the bottom two edges are both $\ket{\psi_0}$, whereas
  the boundaries in region $A$ are ``inputs'' that determine the
  matrix element of the reduced density matrix.
}\label{fig:reduced-density-matrix}
\end{figure}

We now have all of the tools to understand how to compute the
R\'{e}nyi entanglement entropy as a function of time after the quench.
First note that it should now be clear how to compute the reduced
density matrix \eqref{dm}:
\begin{calc}
\bra{a_2}\hat{\rho}_A\ket{a_1} &= \bra{a_2}\Tr_{B}\hat{\rho}\ket{a_1} \\
   &= \int_B Db\,N_\epsilon 
    Z\big(-iT-\epsilon, -iT+\epsilon; \psi_0; \varphi_1= \{a_1,b\}, \varphi_2 =\{a_2, b\}\big)\\
   &\equiv N_\epsilon Z_A\big(-iT-\epsilon, -iT+\epsilon; \psi_0; a_1, a_2\big).
\end{calc}
Here we indicate a field taking values $a$ on $A$ and $b$ on $B$ by
$\{a,b\}$.  We can compute this quantity from the same path integral
in~\eqref{eq:euc-path-integral} with altered boundary conditions at
$t=0$. In the region $B$, we now demand that $\phi$ be continuous at
$t=0$. We started with a full cut, which we sew together in region $B$.
The boundary conditions on the remainder determine the matrix element
computed. This manifold is pictured in
Figure~\ref{fig:reduced-density-matrix}.

Now, to compute $\Tr_{A} \hat{\rho}_A^n$ we start by inserting $\int
Da\ket{a}\bra{a}$ in between each $\hat{\rho}_A$ and then
perform the trace in the $\ket{a}$ basis. This becomes $n$ distinct
copies of the above path integral with appropriate integrals over the
$a_i$:
\begin{calc}
\label{eq:tr-rho-n}
\Tr \hat{\rho}_A^n
 &= \int D a_0 \int D a_1 \cdots\int D a_{n-1}\ 
   \bra{a_0}\hat{\rho}_A\ket{a_{n-1}}\cdots
   \bra{a_2}\hat{\rho}_A\ket{a_1}\bra{a_1}\hat{\rho}_A\ket{a_0}\\
 &= \int D a_0\cdots\int D a_{n-1} (N_\epsilon)^n\ 
    Z_A(\tau_0, \tau_f; \psi_0; a_{n-1}, a_0)\cdots
    Z_A(\tau_0, \tau_f; \psi_0; a_0, a_1),
\end{calc}
where $\tau_0$ and $\tau_f$ get analytically continued as described.
One can then put all of the pieces into a single path integral over
$n$ replicas, with $n$-twisted boundary conditions in region $A$ connecting the replicas and
singly-twisted boundary conditions outside of $A$.

\subsection{Entanglement entropy in the D1D5 CFT}

Let us apply the above general discussion to the matter at hand. We
need to compute the twisted path integral described above. We can
rewrite the path integral as a correlator of local twist operators
that induce the appropriate monodromy, and then compute the correlator
using techniques in~\cite{lunin_correlation_2001}. 
Let us note
that there is an extra layer of obfuscation beyond
computations in other CFTs since our quench involves a distinct twist
operator that is part of the physical spectrum of the CFT.

We prepare the state $\ket{\psi_0}$ by starting with the vacuum at
$\tau=-\infty$, evolving forward to $\tau_0$ where we insert our
quench $\widehat{O}(w_0)$ from \eqref{eq:def-op}. To compute the R\'{e}nyi entropy, we need
$n$ replicas of $\ket{\psi_0}$. The trace in \eqref{eq:tr-rho-n} is then proportional to the
four-point function
\begin{equation}\label{eq:full-4-pt}
  W_n(\tau_0, \tau_f; \theta_1, \theta_2) = \left\langle[O^\dg(w_f)]^n\,\sigma_n(w_2)\sigma_n(w_1)\,[O(w_0)]^n\right\rangle,
\end{equation}
where
\begin{equation}
w_0 = \tau_0\qquad w_f = \tau_f \qquad w_1 = i\theta_1\qquad w_2 = i\theta_2.
\end{equation}
Since each of the $O$'s has a 2-twist it becomes necessary to clarify
the branching structure of the correlator. Let us specify the indices
of the twist fields involved in the correlator:
\begin{equation}\label{eq:indices}
[\sigma_2]^n = \sigma_{(12)}\sigma_{({3}{4})}\cdots\sigma_{({2n-1},{2n})}\qquad
\sigma_n = \sigma_{(1{3}{5}\dots{2n-1})}.
\end{equation}
The indices are labels for the $2n$ sheets involved in the correlator,
and we use the parenthetical notation for single cycles of $S_n$. Note
that the bare twist $\sigma_n$ introduced by the replica trick twists
one copy from each pair of twisted replicas. This fixes the topology
of the correlator. 

We can now write the R\'{e}nyi entanglement entropy as
\begin{equation}
S_n(T, \theta_1, \theta_2) =-\frac{1}{1-n}\log \left(
\frac{W_n(-iT-\epsilon, -iT+\epsilon; \theta_1, \theta_2)}{[W_1(-iT-\epsilon, -iT+\epsilon)]^n}\right).
\end{equation}
Note that $\sigma_1$ is the identity operator and so there is no need
to specify the $\theta_1$ and $\theta_2$ for $W_1$. 
Also note that any normalization issues from defining
$\ket{\psi_0}$ in terms of the local operator $O(w_0)$ cancel out
between the numerator and denominator.

\section{The four-point function}
\label{sec:4pt}

The four-point function in Equation~\eqref{eq:full-4-pt} factorizes
into a four-point function of bare twist operators that we compute by
mapping to a covering space and a correlator of insertions in the
covering space.  

We first compute the correlator of the bare twists, and then treat the
non-twist supercharge insertions that appear in the covering space. We
map the correlator in Equation~\eqref{eq:full-4-pt} to the plane via
the exponential map~\eqref{eq:exp-map}. We will then treat the
associated Jacobian factors in \S\ref{sec:cyl-to-plane}.

\subsection{The twist correlator}
\label{sec:twist-corr}

Let us begin, then, with just the twist part of the correlator
\begin{equation}\label{eq:bare-4-pt-corr}
\braket{[\sigma_2(z_f)]^n {\sigma}_n(z_3){\sigma}_n(z_2)[\sigma_2(z_0)]^n}.
\end{equation}
Note that this part of the correlator applies to any CFT with two
copies that are suddenly joined by $\sigma_2$. Thus when discussing
the bare twist results, we keep $c$ the central charge of a single
copy.  The $\mathrm{SL}(2,\co)$ symmetry determines the form of the 4-pt
function up to an arbitrary function of the cross-ratio. Therefore, we
can compute the 4-pt function
\begin{equation}\label{eq:4-pt-01inf}
\mathcal{F}_n(u) = \braket{[\sigma_2(\infty)]^n{\sigma}_n(u){\sigma}_n(1)[\sigma_2(0)]^n},
\end{equation}
and then find the 4-pt function of interest in
Equation~\eqref{eq:bare-4-pt-corr}.

To compute the four-point function we need to find a map to the
covering space, and then compute the Liouville action associated with
the map~\cite{lunin_correlation_2001}. Fortunately, Appendix~D
of~\cite{avery_using_2010} gives an explicit formula for spherical
genus correlation functions of $S_N$-twist operators as a function of
the coefficients of the map. Once we find the map, we can make use of
the formula to avoid computing the Liouville action directly.

Let us now list the properties the map $z=z(t)$ from the $z$-plane to
the $t$-plane must have, as determined by the index structure shown in
Equation~\eqref{eq:indices}. First, one can show from the
Riemann--Hurwitz formula that the covering space must have spherical
genus:
\begin{equation}
g = \frac{1}{2}\sum_i r_i - s + 1 = \frac{1}{2}[1\cdot n +(n-1)+(n-1)+1\cdot n] - (2n) + 1 = 0,
\end{equation}
where $r_i$ is the ramification of the $i$th point with nontrivial
monodromy and $s$ is the total number of sheets (or indices) involved.
Second, the map must have monodromy at $z=0$, $1$, $u$, and $\infty$
appropriate for their respective twist operators. For example, the
point $z=0$ must have $n$ images in the covering space, each with
monodromy 2. Third, generic points in the $z$-plane should have $2n$
distinct images in the $t$-plane.  Thus, we are looking for a
meromorphic function $z(t)$ with the following local
properties:\footnote{Let us note that in an unfortunate notational
  choice the stars do \emph{not} indicate complex conjugation. They
  are merely labels.}
\begin{equation}\begin{aligned}
z &\approx a^*_j (t-t^*_j)^2 & z &\approx 0, t\approx t^*_j\qquad j = 0, \dots, n-1\\
z - 1 &\approx a_1 t^n & z &\approx 1, t\approx 0\\
z - u &\approx a_u (t-1)^n & z&\approx u, t\approx 1\\
z &\approx b_0 t^2 & z&\to \infty, t\to t^\infty_0 = \infty\\
z &\approx \frac{b_j}{(t-t_j^\infty)^2} & z&\to\infty, t\approx t^\infty_j\qquad j=1, \dots, n-1,
\end{aligned}\end{equation}
where the $a$'s and $b$'s are coefficients that are determined from
the map and the $t^*_j$ and $t_j^\infty$ are the various preimages of $z=0$ and $z=\infty$, respectively. We have fixed the points $0$, $1$, and $\infty$ in the
$t$-plane. The remaining points in the $t$-plane must be determined
from the map.

A rational function that satisfies the above properties is given by
\begin{equation}
z = \left[\frac{A t^n - (t-1)^n}{t^n - (t-1)^n}\right]^2,
\end{equation}
where
\begin{equation}
u = A^2.
\end{equation}
Note that 
\begin{equation}
t^*_j = \frac{1}{1-A^\frac{1}{n}e^{i\frac{2\pi j}{n}}}\qquad
t^\infty_j = \frac{1}{1-e^{i\frac{2\pi j}{n}}},
\end{equation}
and we can write
\begin{equation}
\der{z}{t} = \frac{2 n (A-1)^2}{[t^n-(t-1)^n]^3}t^{n-1}(t-1)^{n-1}\prod_{j=0}^{n-1}(t-t^*_j)
\end{equation}
and
\begin{equation}\label{eq:prod-ids}
 t^n - (t-1)^n =n \prod_{j=1}^{n-1} (t-t^\infty_j)\qquad
 A t^n - (t-1)^n = (A-1)\prod_{j=0}^{n-1}(t-t_j^*).
\end{equation}
This allows us to write the map in a form more conducive to finding
the $a$'s and $b$'s,
\begin{equation}
z = \frac{(A-1)^2}{n^2}\frac{\prod_{j=0}^{n-1}(t-t^*_j)^2}{\prod_{j=1}^{n-1}(t-t_j^\infty)^2}.
\end{equation}
The coefficients can be written as
\begin{subequations}
\begin{align}
a_1 &= 2(-1)^{n+1}(A-1)\\
a_u &= 2A(A-1)\\
a^*_k &= \frac{n (A-1)^2}{[(t^*_k)^n-(t^*_k-1)^n]^3}(t^*_k)^{n-1}
        (t^*_k-1)^{n-1}\prod_{j=0, j\neq k}^{n-1}(t^*_k-t^*_j)\\
b_0 &= \frac{(A-1)^2}{n^2}\\
b_k &= \frac{(A-1)^2}{n^2}\left[\frac{\prod_{j=0}^{n-1}(t^\infty_k - t^*_j)}
                   {\prod_{j=1,j\neq k}^{n-1}(t^\infty_k-t^\infty_j)}\right]^2.
\end{align}
\end{subequations}
We can plug these into the formula from~\cite{avery_using_2010} to
find $\mathcal{F}_n(u)$,
\begin{equation}\label{eq:avery-form}
\mathcal{F}_n(u) = 
\left(\prod_{i=1}^M p_i^{-\frac{c}{12}(p_i+1)}\right)
\left(\prod_{j=0}^{N-1} q_j^{\frac{c}{12}(q_j-1)}\right)
\left(\prod_{i=1}^M|a_i|^{-\frac{c}{12}\frac{p_i-1}{p_i}}\right)
\left(\prod_{j=0}^{F-1}|b_j|^{-\frac{c}{12}\frac{q_j+1}{q_j}}\right) 
     |b_0|^\frac{c}{6} q_0^\frac{c}{6},
\end{equation}
where
\begin{equation}
p^*_j \equiv 2\qquad p_1 =p_w = n\qquad q_j \equiv 2,
\end{equation}
and $c$ is the central charge of a \emph{single} copy of the CFT.
(Remember that $c=6$ for the D1D5 CFT.) All of the various
coefficients and products can be checked by appropriate Laurent
expansions and by judicious use of the
identities~\eqref{eq:prod-ids}.\footnote{One needs to plug in with
  specific values of $t$ and take derivatives to get the desired
  results.} The final result takes a very simple form
\begin{calc}\label{eq:4-pt-01winf}
\mathcal{F}_n(A) &= 
  \left(2^{-\frac{c}{4}n} n^{-\frac{c}{6}(n+1)}\right)
  \left( 2^{\frac{c}{12}n}\right)
  \left(|4 A(A-1)^2|^{-\frac{c}{12}\frac{n-1}{n}}\left|n^{2n}\frac{A^{2n-2}}{(A-1)^{2n-4}}\right|^{-\frac{c}{24}}\right) \\
  &\quad \times \left|\frac{(A-1)^{2n}}{n^{2n+4}}\right|^{-\frac{c}{8}}
  \left|\frac{(A-1)^2}{n^2}\right|^\frac{c}{6}
  2^{\frac{c}{6}}\\
 &= \left|4 A (A-1)^2\right|^{-\frac{c}{12}(n-\frac{1}{n})},
\end{calc}
where recall that $A=\sqrt{u}$. As usual, this statement is ambiguous
since the square root has two branches. For instance, as one moves $u$
to $z=1$, $A$ approaches either $+1$ or $-1$; the amplitude is
zero in one case and not in the other. The location of the branch cut
plays a crucial role in getting the correct answer, so let us take a
moment to discuss its physical origin.

There are two sets of branch cuts associated with the correlator in
Equation~\eqref{eq:4-pt-01inf}. There are the
$(12)({3}{4})\cdots$ branch cuts which extend from the origin
to infinity, and there is the $(1{3}{5}\cdots)$ branch cut
that connects the two $\sigma_n$'s. From this structure, we see that
each of the two $\sigma_n$'s can be on one of two branches.
Physically, then, when $u$ approaches $1$ the two $\sigma_n$'s could
be on the same branch and colliding, or they could be on two separate
branches and widely separated. This explains the physical origin of
the square root.

\subsection{The non-twist insertions}

We now include the effect of the non-twist operators. Each application
of $G_{-\frac{1}{2}}$ brings in a Jacobian factor when mapped to the
covering space. The chiral primary $\sigma_2^{++}$ inserts an
appropriately normalized spin field $S^{++}$ in the covering
space~\cite{lunin_three-point_2002}; all other Jacobian factors associated with
the chiral primary twist operator are taken care of by the Liouville
action and the operator normalization.

Thus for each operator inserted at the origin, we write (this is the
holomorphic part)
\begin{calc}
\sqrt{2}\left(G^-_{\dot{A},-\frac{1}{2}}\sigma_2^{+}(0)\right) &= 
  \sqrt{2}\oint_0 \frac{\drm z}{2\pi i}G^-_{\dot{A}}(z) \sigma_2^+(0)\\
 &\rightarrow \sqrt{2}\oint_{t^*}\frac{\drm t}{2\pi i} 
   \left(\der{z}{t}\right)^{-\frac{1}{2}}G^-_{\dot{A}}(t)\, 
   \left[(a^*)^{-\frac{1}{8}}S^+(t^*)\right]\\
 &=  \frac{1}{(a^*)^{\frac{5}{8}}}\oint_{t^*}\frac{\drm t}{2\pi i} 
   \frac{1}{\sqrt{t-t^*}}\left(\frac{\pd X_{A\dot{A}}(t^*)S^A(t^*)}{\sqrt{t-t^*}} + \dots\right)\\
 &= \frac{1}{(a^*)^\frac{5}{8}}\pd X_{A\dot{A}}(t^*)S^A(t^*).
\end{calc}
The factor of $a^*$ in the square brackets with $S^+$ is a local
normalization that comes with the definition of
$\sigma_2^+$~\cite{lunin_three-point_2002}. We have suppressed the $j$ index on
$a^*$ and $t^*$ in the above. Let us define
\begin{equation}
O_{\dot{A}}(t) = \pd X_{A\dot{A}}(t)S^A(t).
\end{equation}

For the operators inserted at infinity we can write a similar
expression; however, it is helpful to make what we mean by
``infinity'' more precise by making the complex plane into a sphere.
Following~\cite{lunin_correlation_2001}, we cut both the $z$- and
$t$-planes into large discs with all operators in the finite plane
enclosed, and then glue equal-sized discs on top. Then infinity of the
original plane becomes a point centered on this ``second'' disc. The
radius of the $z$-plane discs is $1/\delta$ and the radius of the
$t$-plane discs is $1/\delta'$.

The insertions at infinity become
\begin{equation}
\left((G^\dg)^{\dot{A}}_{-,-\frac{1}{2}}\sigma_{2,+}(\infty)\right)
 \rightarrow \delta^\frac{5}{4}\, b^\frac{5}{8}\, (O^\dg)^{\dot{A}}(t^\infty),
\end{equation}
where again we have suppressed the index $j$ on $b$ and $t^\infty$.
The exception to the above is the $j=0$ insertion at
$t^\infty_0=\infty$. This insertion, gets an additional factor of $\delta'$:
\begin{equation}
\left((G^\dg)^{\dot{A}}_{-,-\frac{1}{2}}\sigma_{2,+}(\infty)\right)
 \rightarrow \frac{\delta^\frac{5}{4}}{{\delta'}^\frac{5}{2}}\, 
 b^\frac{5}{8}\, (O^\dg)^{\dot{A}}(\tilde{t}=0).
\end{equation}

The above equations were written for just the left part of the
operators, but we must also include the right part. Thus, the
contribution from the non-twist insertions may be written as
\begin{multline}
\left|\frac{\prod_j b_j}{\prod_j a^*_j}\right|^\frac{5}{4}
\frac{\delta^{\frac{5}{2}n}}{{\delta'}^5}\\
\times\vev{\no{(O^\dg)^{\dot{A}\dot{B}} (t^\infty_0) (O^\dg)^{\dot{A}\dot{B}} (t^\infty_1)\cdots 
     (O^\dg)^{\dot{A}\dot{B}} (t^\infty_{n-1})}\,\no{
     O_{\dot{A}\dot{B}}(t^*_0)\cdots O_{\dot{A}\dot{B}}(t^*_{n-1})}}\quad (\text{no sum}),
\end{multline}
where the insertion at $t^\infty_0$ should be thought of as at
$t=1/\delta'$ (really at $\tilde{t}=0$); the end result being that the
$\delta'$s cancel out in the limit as $\delta'\to 0$. We can evaluate
the products of the $a^*$s and $b$s as before:
\begin{multline}
\label{eq:fulldefcorr}
\frac{\delta^{\frac{5}{2}n}}{{\delta'}^5} 
\left[\frac{|A-1|^{5(n-1)}}{n^{5(n+1)}|A|^{5\frac{n-1}{2}}}\right]\\
\times\vev{\no{(O^\dg)^{\dot{A}\dot{B}} (\infty) (O^\dg)^{\dot{A}\dot{B}} (t^\infty_1)\cdots 
     (O^\dg)^{\dot{A}\dot{B}} (t^\infty_{n-1})}\,\no{
     O_{\dot{A}\dot{B}}(t^*_0)\cdots O_{\dot{A}\dot{B}}(t^*_{n-1})}}\quad (\text{no sum}).
\end{multline}
The bracketed expression is the Jacobian factor from the map and the
factor of $\delta^{5n/2}$ corresponds to the fact that we have $n$
operators with weight $5/8$ (subtracting off the weight of the bare
twist) in the left and right sectors inserted at infinity in the
$z$-plane. Note that the correlator of bare twist operators has a
factor of $\delta^{3n/2}$ that has been dropped
in~\eqref{eq:avery-form}, since we really want the regularized
correlator.

Since we are dealing with free fields the above correlator is easily
evaluated in terms of Wick contractions.  For instance, for $n=2$ we
have
\begin{calc}
\frac{1}{{\delta'}^5}\vev{\no{(O^\dg)^{\dot{A}\dot{B}} (t^\infty_0) (O^\dg)^{\dot{A}\dot{B}} (t^\infty_1)}\,
     \no{O_{\dot{A}\dot{B}}(t^*_0)O_{\dot{A}\dot{B}}(t^*_{1})}}
 &= \frac{1}{|t^\infty_1 - t_0^*|^5} + \frac{1}{|t^\infty_1 - t_1^*|^5}\\
 &=\left|\frac{1}{2}-\frac{1}{1-\sqrt{A}}\right|^{-5}
  +\left|\frac{1}{2}-\frac{1}{1+\sqrt{A}}\right|^{-5}\\
 &= 2^{5}\frac{|1-\sqrt{A}|^{10}+|1+\sqrt{A}|^{10}}{|1-A|^5}.
\end{calc}

Thus putting all of the contributions together for $n=2$ we find
\begin{calc}\label{eq:F-hat-2}
\widehat{\mathcal{F}}_{2} &= \left|4 A (A-1)^2\right|^{-\frac{3}{4}}\,\cdot\,
\left[\frac{|A-1|^{5}}{2^{15}|A|^{\frac{5}{2}}}\right]\,\cdot\,2^5
\frac{|1-\sqrt{A}|^{10} + |1+\sqrt{A}|^{10}}{|1-A|^5}\\
 &= 2^{-\frac{23}{2}} |A|^{-\frac{13}{4}}
 |A-1|^{-\frac{3}{2}}\left(|1-\sqrt{A}|^{10} + |1+\sqrt{A}|^{10}\right).
\end{calc}
We use this to find the R\'{e}nyi entropy of order 2.

Unfortunately, we could not find a closed-form expression for the
general covering space amplitude suitable for analytic continuation.
It may, however, be written as the sum over all total Wick
contractions:
\begin{multline}
\label{eq:wickn}
\vev{\no{(O^\dg)^{\dot{A}\dot{B}} (\infty) (O^\dg)^{\dot{A}\dot{B}} (t^\infty_1)\cdots 
     (O^\dg)^{\dot{A}\dot{B}} (t^\infty_{n-1})}\,\no{
     O_{\dot{A}\dot{B}}(t^*_0)\cdots O_{\dot{A}\dot{B}}(t^*_{n-1})}}\\
 = \sum_{s\in S_n}\prod_{j=0}^{n-1} |t_{j,s(j)}|^{-5},
\end{multline}
where $t_{j,k} = t^\infty_j-t^*_k$.

\subsection{The four-point function of interest}

We can use~\eqref{eq:4-pt-01winf} to find the 4-point function that we
actually want. The general form of the 4-point function is dictated by
$\mathrm{SL}(2,\co)$ symmetry to be (cf.~\cite{difrancesco_conformal_1999})
\begin{equation}
A_4 = 
\vev{\phi_4(z_4)\phi_3(z_3)\phi_2(z_2)\phi_1(z_1)} = f(\eta, \bar{\eta})
\prod_{i<j} z_{ij}^{\frac{h}{3}-h_i-h_j}\bar{z}_{ij}^{\frac{\bar{h}}{3}-\bar{h}_i-\bar{h}_j}
   \qquad \eta = \frac{z_{12}z_{34}}{z_{13}z_{24}},
\end{equation}
where $z_{ij} = z_{i}-z_{j}$ and $f(\eta, \bar{\eta})$ is a function that is completely
undetermined by $\mathrm{SL}(2,\co)$ symmetry. We can compute the 4-point
function with points $0$, $1$, $u$, and $\infty$, and determine $f$
and therefore the general 4-point function.

\subsubsection{The four-point function of bare twists}

We separately compute the entanglement entropy of the bare twist
operator $\sigma_2$ and of the full deformation operator.

In our case the operators are left--right symmetric and therefore $h_i
= \bar{h}_i$, and 
\begin{equation}\label{eq:A4}
A_4 = f(\eta) \prod_{i<j} |z_{ij}|^{2(\frac{h}{3}-h_i-h_j)}.
\end{equation}
We know that the conformal scaling dimensions for the bare twists are \cite{knizhnik_analytic_1987, calabrese_entanglement_2004}
\begin{equation}
h_1 = h_4 = n\frac{c}{24}\left(2-\frac{1}{2}\right)\qquad
h_2 = h_3 = \frac{c}{24}\left(n-\frac{1}{n}\right),
\end{equation}
Above, we computed the correlator with
\begin{equation}
z_1 = 0 \qquad z_4 = \infty\qquad z_2 = 1\qquad z_3 = u.
\end{equation}
For this case, we have
\begin{equation}
\eta = \frac{1}{u}
\end{equation}
and thus the 4-point function takes the form
\begin{equation}
A_4 = 
|\infty|^{-4h_4}\,
        \left[f(\eta)|1-u|^{2(\frac{h}{3}-2h_2)}|u|^{2(\frac{h}{3} -h_1-h_2)}\right],
\end{equation}
where we regulated the factor of ``$\infty$'' by putting the CFT on a
disc (see above). We are interested in the finite part.

We can then write the general 4-pt function in terms of the one
we computed as (we use the fact that $h_1=h_4$ and $h_2=h_3$)
\begin{equation}\label{eq:gen-4-pt}
A_4 = \mathcal{F}_n(u) \left|\frac{\eta}{1-\eta}z_{23}\right|^{-4h_2}|z_{0f}|^{-4h_1}
\qquad u = \frac{1}{\eta}.
\end{equation}
Plugging in with the above weights and with $\mathcal{F}$ from
Equation~\eqref{eq:4-pt-01winf}
\begin{equation}
\vev{\sigma_2(z_f) {\sigma}_n(z_2){\sigma}_n(z_3)\sigma_2(z_0)} =
|z_f - z_0|^{-\frac{c n}{4}}
\left|\frac{(1+\sqrt{\eta})^4}{16\,\eta\,(z_2 - z_3)^4}\right|^{\frac{c}{24}(n-\frac{1}{n})},
\end{equation}
where recall
\begin{equation}
\eta = \frac{(z_0 - z_2)(z_3-z_f)}{(z_0-z_3)(z_2 - z_f)}\qquad
1-\eta = \frac{(z_0-z_f)(z_2-z_3)}{(z_0-z_3)(z_2 - z_f)}.
\end{equation}
Note that the amplitude is invariant under interchange of $z_2$ and
$z_3$ or $z_0$ and $z_f$.

\subsubsection{The four-point function with the full deformation operator for $n=2$}

For the full deformation operator the weights are given by
\begin{equation}
h_1=h_4 =n \qquad h_2=h_3 = \frac{1}{4}\left(n-\frac{1}{n}\right).
\end{equation}
Plugging in as before, this gives
\begin{equation}
\vev{[\widehat{O}^\dg(z_f)]^n\sigma_n(z_3)\sigma_n(z_2)[\widehat{O}(z_0)]^n}
 = \widehat{\mathcal{F}}_n(u) \left|\frac{\eta}{1-\eta}z_{23}\right|^{-(n-\frac{1}{n})}|z_{0f}|^{-4n},
\end{equation}
where as before we should replace $u$ with $1/\eta$.
 
For the case $n=2$, we can plug in with $\widehat{\mathcal{F}}_2$ from
Equation~\eqref{eq:F-hat-2}:
\begin{equation}
\vev{[\widehat{O}^\dg(z_f)]^2\sigma_2(z_3)\sigma_2(z_2)[\widehat{O}(z_0)]^2}
 = 2^{-\frac{23}{2}}|z_{23}|^{-\frac{3}{2}}|z_{0f}|^{-8}
   \left|\frac{1+\sqrt{\eta}}{\eta^{\frac{1}{4}}}\right|^{\frac{3}{2}}
     \frac{|1-\eta^{\frac{1}{4}}|^{10}+|1+\eta^{\frac{1}{4}}|^{10}}{|\eta|^\frac{5}{4}}.
\end{equation}
We have carefully written the above expression so that the
$\eta\mapsto 1/\eta$ symmetry is manifest. This symmetry comes from
the exchange symmetry $z_2 \leftrightarrow z_3$ or $z_0
\leftrightarrow z_f$.

\subsection{From the cylinder to the plane}
\label{sec:cyl-to-plane}

The physics of the D1D5 system originates on the (Lorentzian)
cylinder, so we should be careful to put in Jacobian factors that
arise in using the exponential map from the cylinder to the plane.
Normalized states are normalized states, so we do not need to worry
about Jacobian factors for the $O_i$ and $O_i^\dg$. We do, however,
need to worry about Jacobian factors from the replica twists.

We started on the plane with dimensionful coordinates $t\in\re$ and $y\in[0, 2\pi R)$
and then introduced Euclidean dimensionless coordinates $\tau$, $\theta$ via \eqref{eq:cyl} that may be
written as a complex coordinate $w = \tau + i\theta$.
Note that $R$ is the radius of the large $S^1$ cycle that the D1s
wrap. Finally, we map to the complex plane using the map in \eqref{eq:exp-map}:
\begin{equation}\begin{aligned}
\label{eq:cyl-map}
z &= e^w = e^{\tau + i \theta} \rightarrow e^{-i\frac{t}{R} + i \frac{y}{R}}\\
\bar{z} &= e^{\bar{w}} = e^{\tau - i\theta}\rightarrow e^{-i\frac{t}{R}-i\frac{y}{R}},
\end{aligned}\end{equation}
where the arrows show how we should analytically continue back to real
time at the end of the calculation.


This is a convenient point in the discussion to put in the appropriate
normalization so that
\begin{equation}
N\vev{O^\dg O}= 1 \Longrightarrow
N = |z_{0f}|^{4h_1}.
\end{equation}
This is in the $z$-plane, but as mentioned above, we can put in the
normalization on the cylinder or on the $z$-plane. One can check that
the normalization ensures that we insert a normalized state, which
means that $W_1=1$.  Note that this factor cancels out the inverse
factor in Equation~\eqref{eq:gen-4-pt}.  Starting from the cylinder,
we can compute the four-point function as
\begin{calc}
W_n &= 
|z_{0f}|^{4h_1}\vev{O_i^\dg{\sigma}_n(w){\sigma}_n(w + i\Delta\theta) O_i}\\
 &= |z_{0f}|^{4h_1}\left|\frac{z}{R}\right|^{4\Delta_n}
\vev{O_i^\dg{\sigma}_n(z){\sigma}_n(z e^{i\Delta\theta}) O_i}\\
 &= |z_{0f}|^{4h_1}\left|\frac{z}{R}\right|^{\frac{c}{6}(n-\frac{1}{n})}
\vev{O_i^\dg{\sigma}_n(z){\sigma}_n(z e^{i\Delta\theta}) O_i}.
\end{calc}
This normalization ensures that $W_{1} = 1$ and this will let us
easily compute the entanglement entropies. Note that while the
Jacobian factor is unity for $n=1$, it still gives a nontrivial
contribution to the von Neumann entropy defined as the limit as $n\to
1$.

\section{The entropy}
\label{sec:entropy}

We now have all of the pieces to discuss the entropy. Let us first
treat the R\'{e}nyi entanglement entropy of the bare twist operator:
\begin{calc}\label{eq:Snbare}
S_n &= \frac{1}{1-n}\log\frac{W_n}{W_1^n}\\
    &= \frac{c}{24}\frac{n+1}{n}
    \log\left[R^4\frac{|z_2-z_3|^4}{|z_2|^2|z_3|^2}
    \left|\frac{16\eta}{(1+\sqrt{\eta})^4}\right|\right].
\end{calc}
Since we have carefully normalized the $W_n$ so that $W_1 = 1$, the
von Neumann entropy is 
\begin{equation}\label{eq:svn-final}
S_{\text{vN}} = \lim_{n\to 1}S_n 
  = \frac{c}{12}\log\left[R^4\frac{|z_2-z_3|^4}{|z_2|^2|z_3|^2}
                          \left|\frac{16\eta}{(1+\sqrt{\eta})^4}\right|\right].
\end{equation}
The limit is trivial only because we already canceled out the
$1-n$ in the denominator.  We illustrate
various properties of this formula in \S\ref{sec:sample}.

Let us also write down the $n=2$ R\'{e}nyi entropy for the full
deformation operator. This is given by
\begin{equation}
\label{eq:S2}
S_2 = -\log W_2 = -\log\left[2^{-\frac{23}{2}} R^{-\frac{3}{2}}
\frac{|z_2|^{\frac{3}{4}}|z_3|^\frac{3}{4}}{|z_{23}|^\frac{3}{2}}
\left|\frac{1+\sqrt{\eta}}{\eta^\frac{1}{4}}\right|^{\frac{3}{2}}
\frac{|1-\eta^{\frac{1}{4}}|^{10}+|1+\eta^{\frac{1}{4}}|^{10}}{|\eta|^\frac{5}{4}}
\right].
\end{equation}
Recall that the R\'{e}nyi entropy gives a lower bound for the von
Neumann entropy and that it vanishes if and only if the reduced
density matrix is that of a pure state.



\subsection{Entanglement of the vacuum}

It is useful to subtract off the contribution to the entanglement
entropy from the vacuum, which we compute here. We want to compute
this as a function of the physical cylinder coordinates, so we include
Jacobian factors for the map $z \rightarrow y$, while keeping the
convenient variables $z = e^{iy/R}$. From the above, we have
\begin{calc}\label{eq:Snvac}
S_n^{\text{vac}} &= \frac{1}{1-n}\log\left[\left|\frac{z}{R}\right|^{\frac{c}{6}(n-\frac{1}{n})}
\vev{{\sigma}_n(z){\sigma}_n(z e^{i\theta})}\right]\\
  &= \frac{1}{1-n}\log\left[\left|\frac{1}{R}\frac{z_2}{z_2-z_3}\right|^{\frac{c}{6}(n-\frac{1}{n})}\right]\\
  &= \frac{c}{6}\frac{n+1}{n}\log R \left|\frac{z_2-z_3}{z_2}\right|.
\end{calc}
The limit as $n\to 1$ yields the von Neumann entropy
\begin{calc}
S_{\text{vN}}^{\text{vac}} &= \frac{c}{3}\log R \left|\frac{z_2-z_3}{z_{2}}\right|.
\end{calc}
Since $z_3 = e^{i\theta}z_2$ we have 
\begin{equation}
\left|\frac{z_3-z_2}{z_2}\right| = \left|\frac{z_3}{z_2} - 1\right| = \left|e^{i\theta}-1\right|
   = 2|\sin\tfrac{\theta}{2}|,
\end{equation}
and so
\begin{equation}
\label{vac}
	S_{\text{vN}}^{\text{vac}} = \frac{c}{3} \log\left[ \frac{L}{\pi} \sin\left( \frac{\pi l}{L}\right)\right],
\end{equation}
where we have written the above expression in terms of $l = |y_{2} -
y_{1}|$ and $L = 2\pi R$ to show agreement with
\cite{holzhey_geometric_1994, calabrese_entanglement_2004} for a CFT
on a cylinder.

Actually, in the above we have suppressed an ultraviolet cutoff $a$ in
terms of which we measure the physical lengths $l$ and $L$. So the
correct expression is \eqref{vac} with $l$ and $L$ replaced by $l/a$
and $L/a$. There is a logarithmic divergence of
$S_{\text{vN}}^{\text{vac}}$ as $a \rightarrow 0$ as noted in early
investigations \cite{holzhey_geometric_1994, srednicki_entropy_1993}.
One can either treat the expression as an asymptotic result for a
small but finite $a$, as would be appropriate for studying lattice
theories (e.g., in condensed matter theory), or find ``renormalized''
entropy \emph{differences} between various states in the theory. In
some cases these differences are finite in the $a \rightarrow 0$ limit
\cite {holzhey_geometric_1994}.

Here we follow the latter approach (although, as it turns out, our
entropy differences are still divergent), since we are interested in
the extra entanglement added by the quenching process: the
entanglement entropy increase, $\Delta S$. After subtracting off the
entanglement of the vacuum we get for the insertions of the bare twist
operators
\begin{equation}
\label{eq:svn-rel}
\Delta S_{\text{vN}} = \frac{c}{12}\log\left|\frac{16\eta}{(1+\sqrt{\eta})^4}\right|
\end{equation}
and similarly for the full deformation operators:
\beq
\label{eq:s2-rel}
\Delta S_2 = \log\left[2^{\frac{23}{2}} 
\left|\frac{\eta^\frac{1}{4}}{1+\sqrt{\eta}}\right|^{\frac{3}{2}}
\frac{|\eta|^\frac{5}{4}}{|1-\eta^{\frac{1}{4}}|^{10}+|1+\eta^{\frac{1}{4}}|^{10}}
\right].
\eeq

\section{Properties}
\label{sec:sample}

In this section we include plots of the entropies $\Delta
S_{\text{vN}}$ and $\Delta S_{2}$ for a variety of space and time
intervals and examine some of their properties. We discuss the details
of the UV cutoff and the choices of branch cuts needed to compute with
the entropy formulas we have given.
We start with the
entanglement entropy introduced by the bare twist operator, then
consider the full deformation operator.
	
\subsection{The entanglement from a bare twist}

Our formula \eqref{eq:svn-rel} can now be computed for any time and
choice of interval $[\theta_{1},\theta_{2}]$. We recall that the
quench is located at $\theta = 0$ and $t = 0$. To actually compute
from \eqref{eq:svn-rel} we must choose $c$, $\epsilon$ (as described
in \S \ref{sec:replica}), and a prescription for the square roots (as
mentioned at the end of \S \ref{sec:twist-corr}). In this subsection,
we choose $c = 1$.\footnote{The appropriate choice for the D1D5 CFT
  would be $c=6$; however, the $c$-dependence is just an overall
  constant here, and since the bare twist operator is not the
  deformation operator, we may as well just set $c=1$.} The
entropy increase $\Delta S$ diverges as $\epsilon \rightarrow 0$ for some values of $t$, see \eqref{eq:Speak} below. We can understand this
by thinking of $\epsilon$ as a UV suppression factor or regulator, consistent with its
appearance in the evolution operator $e^{-iH(t-i\epsilon)}$. It is
perhaps not surprising that the local quench has a UV divergence since
it is localized at a point, a phenomenon observed in
\cite{anderson_does_1986}. In a real physical process there should of
course be a finite amount of energy, which we can treat as a finite
$\epsilon$ in our calculation. We would fix the appropriate value of
$\epsilon$ in a full treatment of the physical evolution of the D1D5
CFT, but here we leave it as an unfixed but small parameter. With
finite $\epsilon$ we have
\begin{equation}\begin{split}\label{eq:eta-in-sins}
\eta &= \frac{\sin\left(\frac12 (\frac{t}{R} + \theta_{1}- i\epsilon ) \right) \sin\left(\frac12 (\frac{t}{R} + \theta_{2} + i\epsilon) \right)}{\sin\left(\frac12 (\frac{t}{R} + \theta_{1} + i\epsilon) \right) \sin\left(\frac12 (\frac{t}{R} + \theta_{2} - i\epsilon) \right)}\\
\bar{\eta} &= \frac{\sin\left(\frac12 (\frac{t}{R} - \theta_{1} - i\epsilon) \right) \sin\left(\frac12 (\frac{t}{R} - \theta_{2} + i\epsilon) \right)}{\sin\left(\frac12 (\frac{t}{R} - \theta_{1} + i\epsilon) \right) \sin\left(\frac12 (\frac{t}{R} - \theta_{2} - i\epsilon) \right)}
\end{split},
\end{equation} 
where here $\theta_{i} = y_{i}/R$ and $\epsilon$ should really be
$\epsilon/R$ so that all parameters are dimensionless.  For the
remainder of our discussion, we set $R=1$.  In the above expressions
and in what follows, $\theta$ parametrizes the double circle by
running from $\theta=0$ to $\theta=4\pi$. This explains the factors of
$1/2$ inside the trigonometric functions.

\begin{figure}[ht]
\begin{center}
\includegraphics[width=8cm]{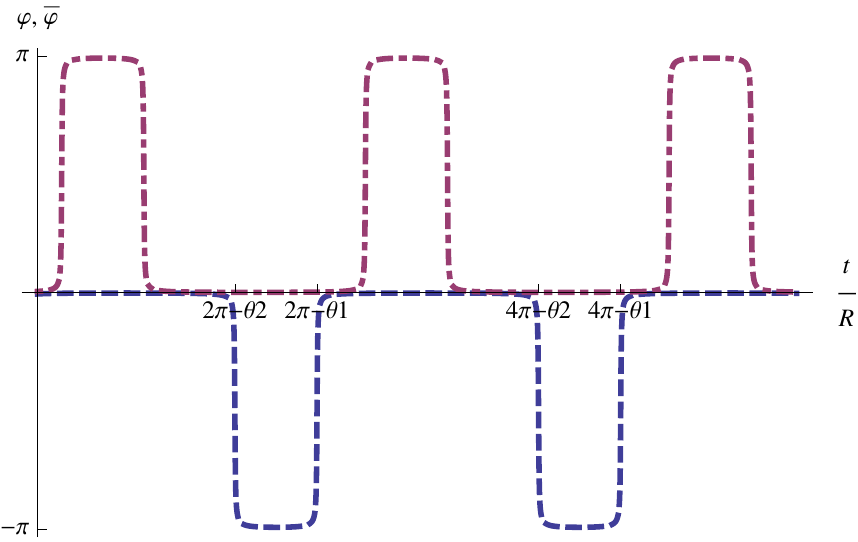}
\end{center}
\caption{Plot of $\varphi$ (blue, dashed) and $\bar{\varphi}$ (red,
  dot-dashed) versus time for $\epsilon = 10^{-2}$, where $\eta = \exp
  2i\varphi$ and $\bar{\eta} = \exp 2i\bar{\varphi}$. Note that
  $\varphi$ and $\bar{\varphi}$ obey the strict inequality
  $-\pi<\varphi<0<\bar{\varphi}<\pi$ for all time. In the limit as
  $\epsilon$ goes to zero, the function converges pointwise to a
  piecewise function saturating the inequalities. Indeed, for
  $\epsilon=10^{-4}$ the plot is (visually) indistinguishable from the
  corresponding piecewise function.}\label{fig:varphi}
\end{figure}

Using sundry trigonometric identities, we can rewrite $\eta$ and
$\bar{\eta}$ in a slightly more useful form
\begin{equation}\label{eq:tanphi}
\eta = e^{2i\varphi}\qquad \tan\varphi 
             = \tilde{\epsilon}\frac{\cot\frac{t+\theta_2}{2}-\cot\frac{t+\theta_1}{2}}
               {1 + \tilde{\epsilon}^2\cot\frac{t+\theta_2}{2}\cot\frac{t+\theta_1}{2}},
\end{equation}
where we have introduced $\tilde{\epsilon} = \tanh(\epsilon/2)$. For
our purposes, it suffices to drop the $\tilde{\epsilon}^2$ term in the
denominator and write
\begin{equation}\label{eq:tanphi-approx}
\tan\varphi \approx 
  \tilde{\epsilon}\left[\cot\tfrac{t+\theta_2}{2}-\cot\tfrac{t+\theta_1}{2}\right] \approx 
   0^{\pm}\Longrightarrow \varphi\approx n\pi\qquad n\in\ints.
\end{equation}
From the above, we conclude that for small $\epsilon$ $\varphi$ is
close to some multiple of $\pi$, and therefore $\eta$ is essentially
unity. This fact could have been read off from~\eqref{eq:eta-in-sins}
directly; the key realization from the above is that the \emph{sign}
of the $0$ depends on time which implies $\varphi$ has some nontrivial
time dependence. If we plot $\varphi$ as function of $t$ for
reasonably small $\epsilon$ and take care with the signs, we get
Figure~\ref{fig:varphi}. Note that $\tan\varphi$ never vanishes
in~\eqref{eq:tanphi}, and therefore $-\pi<\varphi<0$. We have chosen
these particular branches to be consistent with~\eqref{eq:branches}.

All of the time dependence comes from the interaction of this phase
with the square root. We have carefully chosen $\varphi$ and
$\bar{\varphi}$ in Figure~\ref{fig:varphi} so that the correct
branches are
\begin{equation}\label{eq:branches}
\sqrt{\eta} = e^{i\varphi}\qquad \sqrt{\bar{\eta}} = e^{i\bar{\varphi}}\qquad
\eta^\frac{1}{4} = -e^{i\frac{\varphi}{2}}\qquad\bar{\eta}^\frac{1}{4} 
  = -e^{i\frac{\bar{\varphi}}{2}},
\end{equation}
where the fourth roots arise when considering the R\'{e}nyi entropy of
the full deformation operator. Then, we can write the entropy in terms
of $\varphi$ and $\bar{\varphi}$
\begin{equation}
\Delta S_{\text{vN}} = -\frac{c}{6}\log\big( \cos\tfrac12\varphi\big)
                             -\frac{c}{6}\log\big( \cos\tfrac12\bar{\varphi}\big)
\end{equation}
Note that in defining $\varphi$ and $\bar{\varphi}$
from~\eqref{eq:tanphi} there is an ambiguity associated $\tan$;
similarly, in writing the square root of $\eta$ in terms of $\varphi$
and $\bar{\varphi}$ there are two branches one could choose. However,
the above choices are fixed for us by causality: the entanglement
entropy of our interval cannot change from the vacuum value until a
signal traveling at the speed of light from the quench could reach the
interval. The origin of the branch cut is also discussed at the end of
\S\ref{sec:twist-corr}.

Since for small $\epsilon$, $\varphi$ and $\bar{\varphi}$ spend the
vast majority of time near $0$ or $\pm\pi$, let us examine the
limiting behavior of the entropy away from the transitions. This is
where the approximations in Equation~\eqref{eq:tanphi-approx} are
good. Let us define $x$ to be the difference of cotangents,
\begin{equation}\label{eq:def-x}
x = \cot\tfrac{t+\theta_2}{2}-\cot\tfrac{t+\theta_1}{2}.
\end{equation}
Then, for small $\tilde{\epsilon}x$ we can write
\begin{equation}
\cos^2\frac{\varphi}{2} \approx
\begin{cases}
1 - \frac{1}{4}(\tilde{\epsilon}x)^2 + \bigO\big((\tilde{\epsilon}x)^4\big) & x<0 \\
\frac{1}{4}(\tilde{\epsilon}x)^2  + \bigO\big((\tilde{\epsilon} x)^4\big)& x>0
\end{cases},
\end{equation}
and thus the left-moving contribution to the entanglement entropy is
given by
\begin{equation}
S_L = -\log\big(\cos\tfrac{\varphi}{2}\big) \approx
\begin{cases}
\frac{1}{8}(\tilde{\epsilon} x)^2 + \bigO\big((\tilde{\epsilon} x)^4\big) & x<0\\
-\log\frac12\tilde{\epsilon} x + \bigO\big((\tilde{\epsilon} x)^2\big) & x>0
\end{cases}.
\end{equation}
We see that the entanglement entropy (away from transition regions)
either vanishes like $\tilde{\epsilon}^2$ or diverges like
$-\log\tilde{\epsilon}$. The peak value occurs at $t=
-\frac{\theta_1+\theta_2}{2} + 2n\pi$ for integer $n$. We can estimate
the peak value as
\begin{equation}
\label{eq:Speak}
S_L^{\text{peak}} 
 \approx -\log\tilde{\epsilon} + \log\left(\tan\tfrac{\theta_2 - \theta_1}{4}\right)
  + \bigO(\tilde{\epsilon}^2).
\end{equation}
The unbounded growth of $S_L^{\text{peak}}$ as $\epsilon \rightarrow 0$ indicates that the state after the local quench has entangled elements
of arbitrarily high energy, a UV effect already discussed.

The $\epsilon$ dependence of $S_L^{\text{peak}}$ indicates only
partial thermalization, as we now discuss.  We can think of the UV
regulator $\epsilon$ as introducing a temperature $1/2\epsilon$, in
that it corresponds to introducing the operator $e^{-2\epsilon H}$ in
our density matrix given in \eqref{eq:dm}.  We can compare the
$\epsilon$ dependence of $S_L^{\text{peak}}$ with the entanglement
entropy of the interval in the equilibrium mixed state at temperature
$1/2\epsilon$. The limit $\epsilon \rightarrow 0$ corresponds to high
temperatures and in this limit the entropy should be dominated by the
extensive quantity
\begin{equation}
\label{eq:extensive}
S_L^{\text{max}}\bigg|_{\epsilon \rightarrow 0} \sim \frac{\Delta\theta}{\epsilon}.
\end{equation}
This gives the asymptotic behavior of a general CFT of length
$\Delta\theta$ at temperature $1/\epsilon$, which is a regularized
expression for the maximum entropy on that interval. Thus, we see that
while the entanglement entropy we produce diverges, it is
parametrically less than the maximum possible entropy for the
subsystem. In other words, even at peak entanglement, the system is
far from being ``Page-scrambled'' as defined
in~\cite{sekino_fast_2008}, referring to \cite{page_average_1993}.
This is expected since being Page-scrambled would require the reduced
density matrix after the quench to evolve to the maximal-entropy
(infinite-temperature) thermal density matrix, proportional to the
identity operator on $\mathcal{H}_{A}$, but we do not expect a local
quench to lead to a thermal reduced density matrix (at any
temperature). First, this system has decoupled momentum sectors
which can thus be independently thermalized with various
momentum-dependent temperatures \cite{calabrese_quantum_2007}.
Second, the local quench produces coherent sets of noninteracting
particles traveling from the quench point, so there is no mechanism to
scramble their momenta. So thermalization must involve more than the
process we study here, as we discuss further in the Conclusion.
	
Now we can examine some specific calculations of $\Delta
S_{\text{vN}}$, choosing $c = 1$ and $\epsilon = 10^{-4}$. The branch
cuts are chosen as discussed above. The first obvious feature from the
formulas is that the entropy is $2\pi$-periodic, which follows from
the $2\pi$-periodicity of the quenching process; the point where the
two circles join becomes two antipodal points on the length $4\pi R$
circle.

\begin{figure}[h]
\subfloat[][]{
\includegraphics[width=8.5cm]{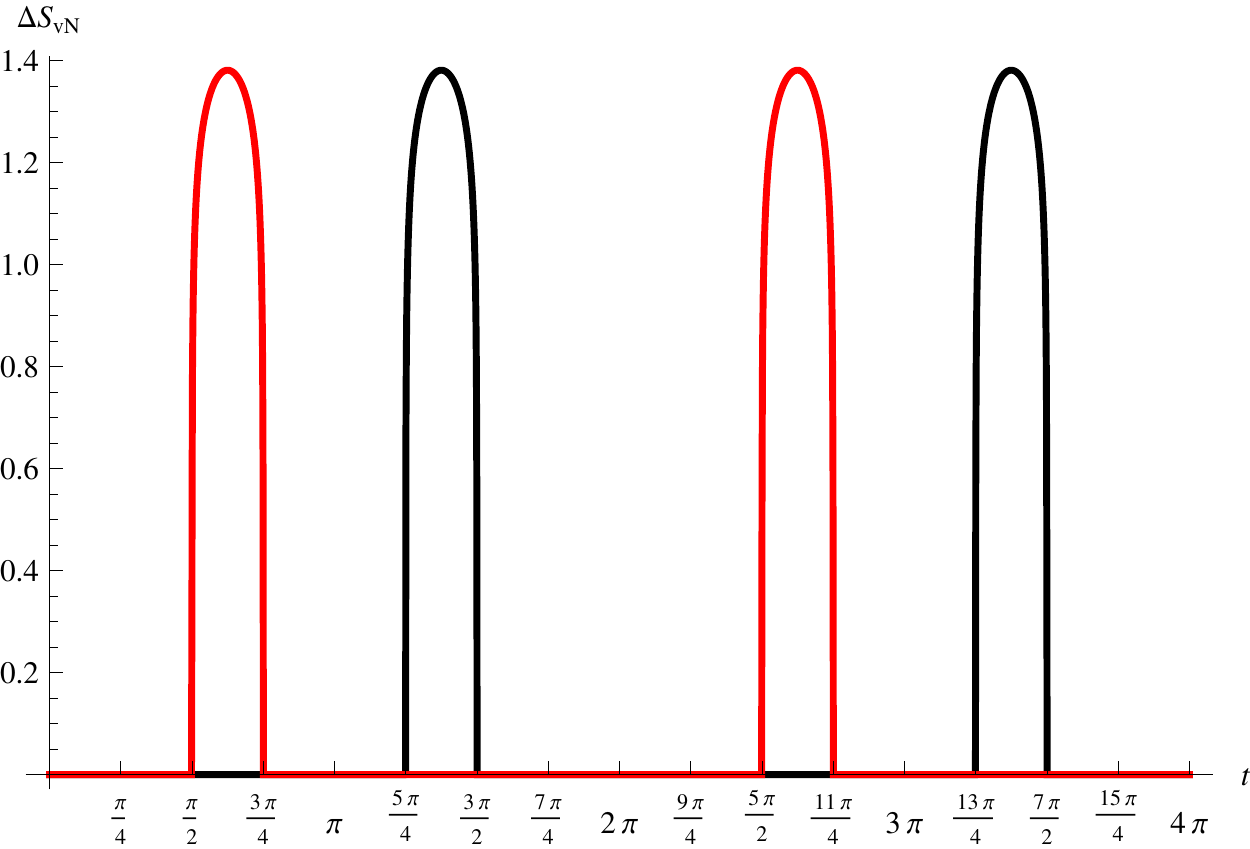}
\label{fig:LRent2}}
\subfloat[][]{
\includegraphics[width=8.5cm]{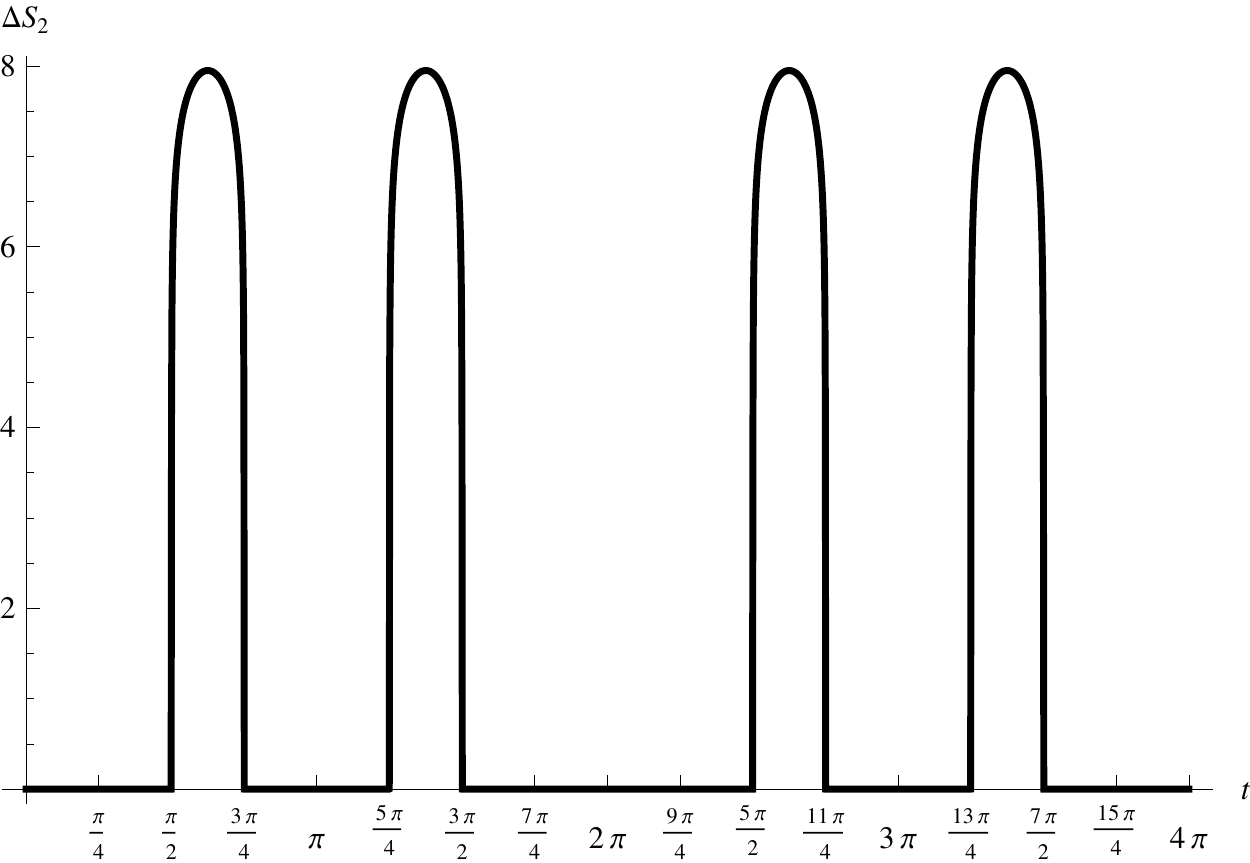}
\label{fig:S2plot}}
\caption{$\Delta S_{\text{vN}}$, \ref{fig:LRent2}, 
and $\Delta S_2$, \ref{fig:S2plot}, 
  for the interval $[\pi/2, 3\pi/4]$. For
  $\Delta S_{\text{vN}}$ the right-moving contribution is shown in
  red, the left-moving in black. For $\Delta S_2$ we cannot
  separate the left and right-moving contributions.}
\end{figure}

\begin{figure}[h]
\begin{center}
\includegraphics[width=5cm]{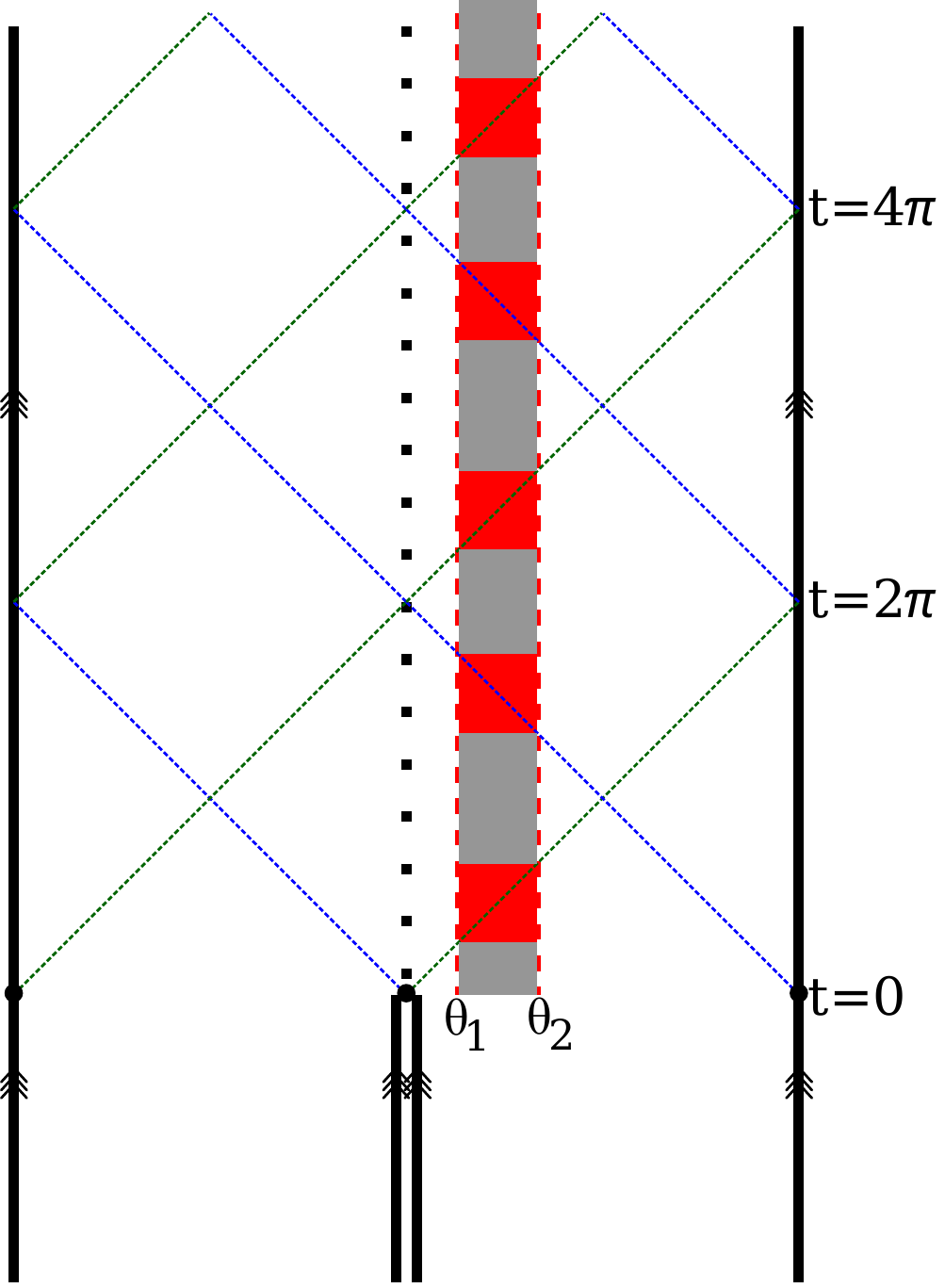}
\end{center}
\caption{Here we show the two circles being joined at $\theta = 0$ and
  $\theta = 2\pi$. Time is increasing up on the figure. The quench
  occurs at $t=0$. Light cones are emanating from the two quench sites.
  The right-moving excitations travel along the green curves, and the
  left-moving excitations along the blue curves. The grey vertical
  strip represents the time evolution of the interval $[\theta_1,
  \theta_2]$, and the red parts of the strip show when we expect
  nonvanishing entanglement from the particle interpretation.}
\label{lightcones}
\end{figure}

In Fig. \ref{fig:LRent2} we choose the interval $[\pi/2, 3\pi/4]$
and see positive entropy at those times when the null world line from
the quench point intersects the interval. We have separated the
contribution from the left-moving and right-moving sectors for
illustrative purposes. With the interval to the right of the quench
point, the entropy can be qualitatively understood in terms of
particles emitted from the quench point and traveling with unit
velocity, a picture first described in
\cite{calabrese_evolution_2005}. The positive entropy comes from the
presence of entangled pairs of (left- or right-moving) particles, one
member of which is inside the interval and the other outside, and so
is traced over. A space-time picture of this is given in
Figure~\ref{lightcones} and we discuss it further in the Conclusion.
$\Delta S_{2}$ behaves the same way, as shown in Figure
\ref{fig:S2plot}. In fact, plots of $\Delta S_{2}$ take almost exactly
the same shapes and share all the qualitative properties of those of
$\Delta S_{\text{vN}}$, so we only include plots of the latter in the
following.

\begin{figure}[h]
\subfloat[][]{
\includegraphics[width=8.5cm]{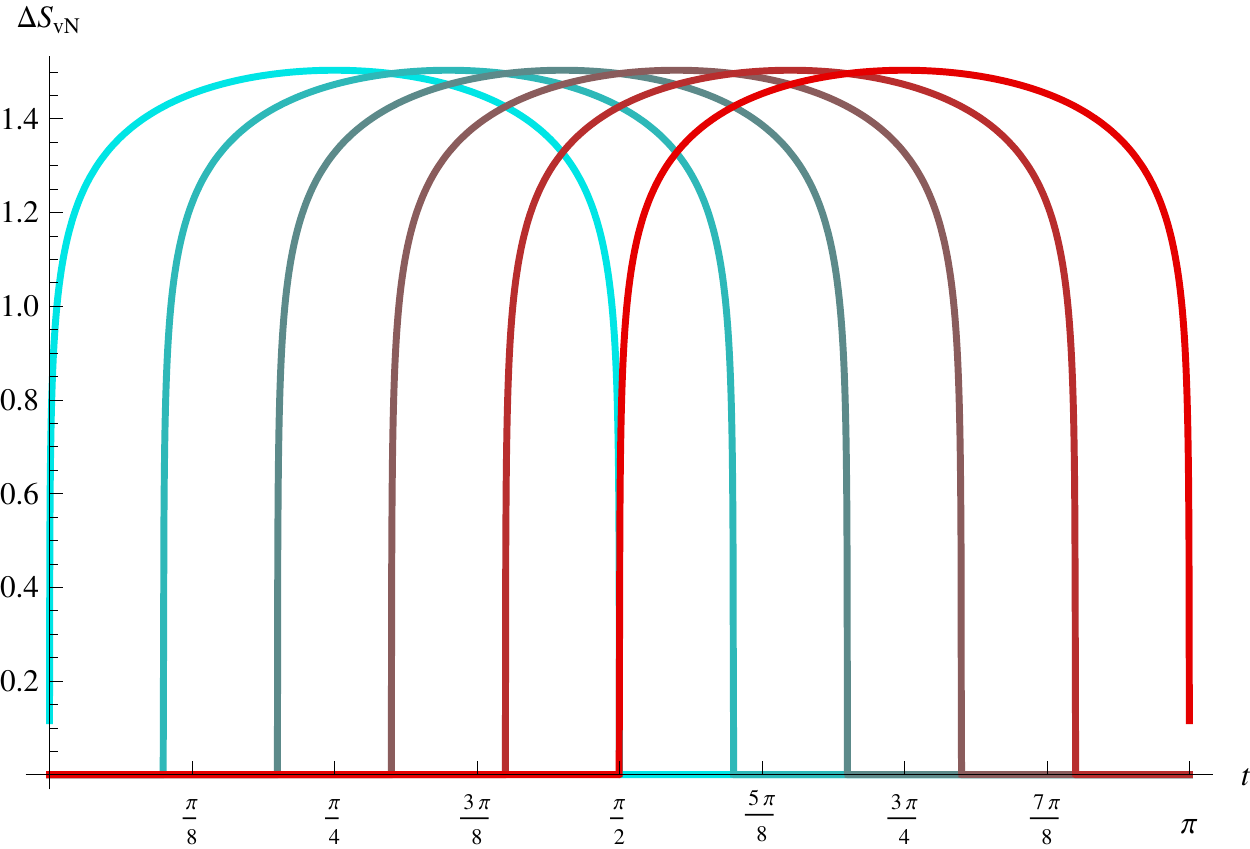}
\label{fig:translate1}}
\subfloat[][]{
\includegraphics[width=8.5cm]{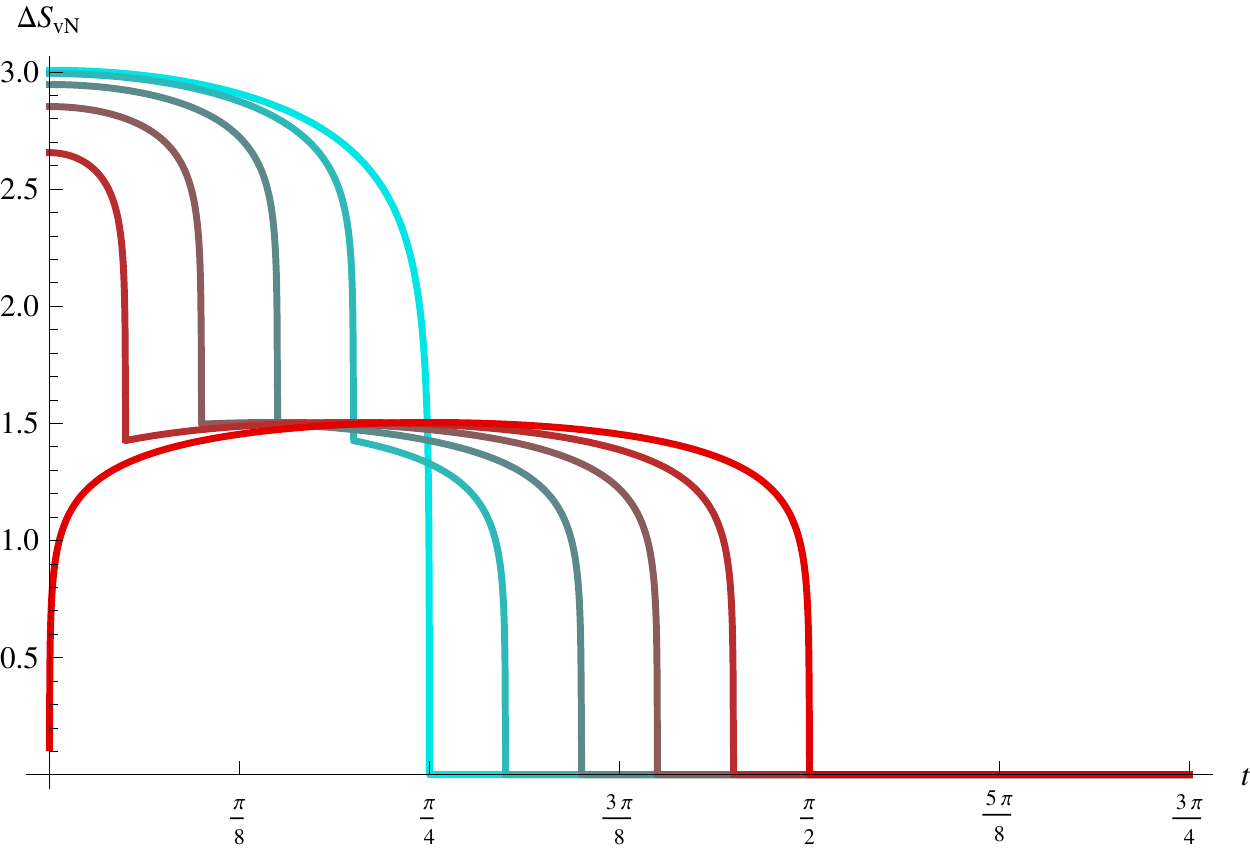}
\label{fig:overlap1}}
\caption{$\Delta S_{\text{vN}}$  for various translated equal-size intervals $[0 + x \pi/10, \pi/2 + x \pi/10]$, \ref{fig:translate1}, and  $[-\pi/4 + x \pi/20, \pi/4 + x \pi/20]$, \ref{fig:overlap1}, with the different colors, going from blue-green to red, showing $x = 0, 1, 2, 3, 4, 5.$}
\end{figure}

\begin{figure}[h]
\begin{center}
\includegraphics[width=8.5cm]{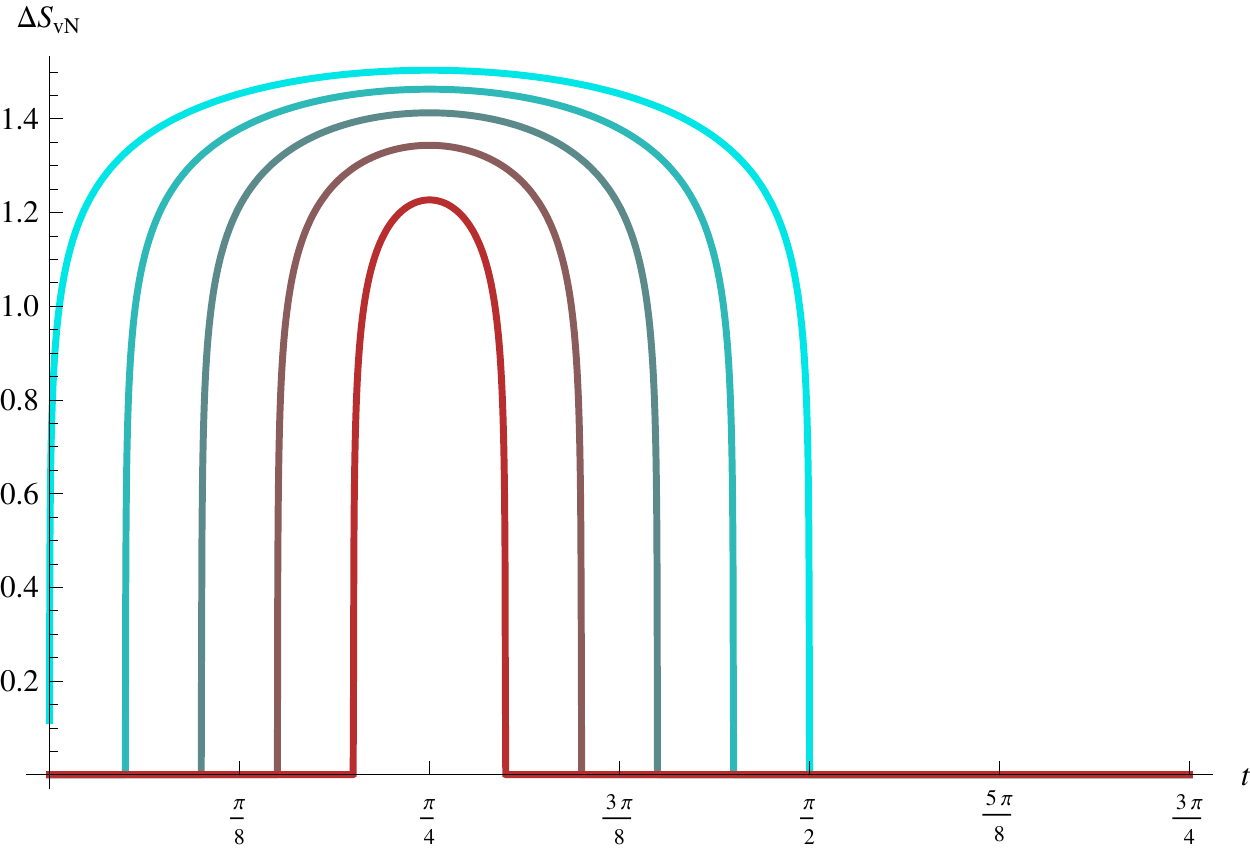}
\end{center}
\caption{$\Delta S_{\text{vN}}$ for intervals of various sizes $[\pi/4 - x \pi/20, \pi/2 + x \pi/20]$, with the different colors, going from red to blue-green, showing $x = 1, 2, 3, 4, 5.$}
\label{fig:size1}
\end{figure}

We can translate the interval toward and away from the quench point,
as seen in Figure~\ref{fig:translate1}, which is seen to have almost no effect on the entropy curve. However, if the quench takes place inside the interval there is a very noticeable effect shown in Figure~\ref{fig:overlap1}, due to overlapping contributions from the left and right-moving sectors. We can examine intervals of
different sizes, as in Figure~\ref{fig:size1}, and see a clear dependence on the size of the interval. The dependence of the peak value on the size can be read from the second term in \eqref{eq:Speak}.



%

%
%
	

%

\subsection{The entanglement from the deformation operator }

Here we examine the second R\'{e}nyi entropy that results from
quenching with the full deformation operator. The expression for $S_2$
in~\eqref{eq:S2} may be written as
\begin{equation}
S_2 = S_2^{\text{vac}} + S_2^{\text{bare}} + S_2^{\text{ins}},
\end{equation}
where $S_2^{\text{vac}}$ is the R\'{e}nyi entropy of the vacuum
in~\eqref{eq:Snvac}, $S_2^{\text{bare}}$ is the additional R\'{e}nyi
entropy added by the bare twist in~\eqref{eq:Snbare}, and
$S_2^{\text{ins}}$ is the new contribution from the supercharge and
spin field in the covering space. We write the three pieces as
\begin{subequations}
\begin{align}
S_2^{\text{vac}} &= \frac{3}{4}\log \left|R^2\frac{z_{23}^2}{z_2z_3}\right|\\
S_2^{\text{bare}} &= \frac{3}{8}\log \frac{16|\eta|}{|1+\sqrt{\eta}|^4}\\
S_2^{\text{ins}} &= -\log\frac{|1-\eta^\frac{1}{4}|^{10}+|1+\eta^\frac{1}{4}|^{10}}
                              {2^{10}|\eta|^\frac{5}{4}}.
\end{align}
\end{subequations}
Using~\eqref{eq:branches}, we can rewrite the last two terms as
\begin{subequations}
\begin{align}
S_2^{\text{bare}} &= -\frac{3}{4}\log\big(\cos\tfrac{\varphi}{2}\big)
                    -\frac{3}{4}\log\big(\cos\tfrac{\bar{\varphi}}{2}\big)\\
S_2^{\text{ins}} &= -\log\left[\cos^5\tfrac{\varphi}{4}\cos^5\tfrac{\bar{\varphi}}{4}
 - \sin^5\tfrac{\varphi}{4}\sin^5\tfrac{\bar{\varphi}}{4}\right],
\end{align}
\end{subequations}
and $\Delta S_{2} = S_2^{\text{bare}} + S_2^{\text{ins}}$. We plot
this for the interval $[\pi/2, 3\pi/4]$ in Figure \ref{fig:S2plot}
and, as mentioned, plots analogous to those in Figures
\ref{fig:translate1}, \ref{fig:overlap1}, and \ref{fig:size1} look
very similar. Let us note that $S_2^{\text{ins}}$ is distinguished
from the entanglement of the vacuum and of the bare twist in that the
left and right contributions do not directly factorize. That being
said, one finds that $S^{\text{ins}}$ still enjoys a superposition
principle with respect to the left and right contributions \emph{as
  long as left- and right-moving excitations do not simultaneously
  contribute to the entropy.} 

To illustrate consider $S_2^\text{ins}$ in the limit of vanishing
$\epsilon$. In particular, the peak value of $S_2^\text{ins}$ when
left or right excitations are separately in the interval is given by
\begin{equation}
\lim_{\epsilon\to 0}S_2^{\text{ins}}\bigg|_{\text{sing. cont.}} = \frac{5}{2}\log 2,
\end{equation}
and is \emph{finite}. When both left and right excitations contribute,
however, $S_2^\text{ins}$ is given by
\begin{equation}
\lim_{\epsilon\to 0}S_2^{\text{ins}}\bigg|_{\text{both cont.}} 
   = 4\log 2,
\end{equation}
which serves as a finite upper bound on $S_2^\text{ins}$; however, it
is \emph{not} simply twice the contribution from left- or right-movers
separately. Aside from the failure of left and right contributions to
factorize, $S_2^{\text{ins}}$ is further distinguished in its
finiteness. In contrast, $S_2^\text{bare}$ represents a huge (actually
divergent) amount of entanglement. This suggests that it is the
twisting part of the deformation operator that predominantly
contributes to thermalization.  This is consistent with the point of
view advocated in~\cite{avery_deforming_2010, avery_excitations_2010,
  avery_intertwining_2011}, which focuses on the effect of the twist
operator.

\section{Conclusion}

We have seen how to analytically compute entanglement entropies for
arbitrary spatial intervals of the D1D5 CFT following a local quench of the
system by an exactly marginal deformation operator, which contains the
2-twist operator and moves the theory toward the supergravity regime
in its moduli space. For the insertions of the bare twist operators we
were able to compute all of the R\'enyi entropies $S_{n}$ and the von
Neumann entropy while for the full deformation operators we were only
able to compute the $S_{n}$ for $n \geq 2$.
We could understand the qualitative behavior of the entropies as arising from pairs of entangled particles generated at the quench event and propagating from there at the speed of light. 

	This process does not lead to thermalization of the system, as we saw in the discussion below \eqref{eq:Speak}, but from the point of view of the interval these particles should appear as thermal radiation in accord with their nonzero entropy. 
The nonzero entropy does not itself, of course, guarantee the radiation is
thermal, but we expect it is for several reasons. First we note
that in general \emph{global} quenches, which can also be understood as producing pairs of entangled particles \cite{calabrese_evolution_2005}, do lead to thermalized systems
\cite{calabrese_quantum_2007, takayanagi_measuring_2010, rigol_alternatives_2011}. In our case
of a local quench, the positive entanglement entropy is apparently due
to the entanglement of pairs of localized excitations, one member of
which is inside the interval and one outside, and so is traced over.
This is a situation familiar from Hawking radiation: it appears
thermal to observers outside the horizon.
	
A toy model illustrates this in more detail. Consider two
harmonic oscillators with operators $\{a,a^{\dagger}\}$ and
$\{b,b^{\dagger}\}$ in the entangled state $|\psi\rangle \propto
e^{\lambda a^{\dagger} b^{\dagger} } |0\rangle$.  Such a squeezed
state is the general result of a Bogolyubov transformation on the
operator algebras and appears in our case as \eqref{eq:sqstate} above.
If we then trace over, say, the $b$ oscillator then the resulting
reduced density matrix is given by 
\beq
\label{eq:reddm}
	\rho_{a} \propto \sum_{n} \abs{\lambda}^{2n} |n\rangle_{a} \langle n |_{a}.
\eeq
Now we can compare to a thermal density matrix for the $a$ oscillator using \eqref{eq:thermst}:
\beq
	\rho_{\text{th}} \propto e^{-\beta H_{a}} \propto \sum_{n} e^{-\beta \omega n} |n\rangle_{a} \langle n |_{a},
\eeq
and we see that $\rho_{a}$
is a thermal density matrix at
inverse temperature $\beta = -\frac1\omega \log\abs{\lambda}^{2}$. It
is straightforward to compute the entropies of the state $\rho_{a}$ as
a function of $\lambda$ (or equivalently $\beta$) but we do not need
the explicit formulas here.

This simple calculation illustrates how thermal density matrices, with
their associated entropies, arise from the process we are considering.
Of course, the trace we perform in the CFT, over the exterior of a
spatial interval, is not directly analogous to this simple case. It
seems to be difficult to perform such a spatial trace directly on the
state after the quench, given by \eqref{eq:sqstate}, which is why we
pursued a technique here that makes extensive use of the powerful
conformal symmetry of the system. It would be interesting to pursue
that direct approach and also to trace over different classes of
subsystems, corresponding to different coarse-grainings.

Ultimately we would like to understand thermalization 
in this
system, with general subsystems characterized by reduced density matrices
of the kind in \eqref{eq:thermst}. Here we have only studied an individual event involving a
small sector of the full theory, which is a
system of $N_{1}$ D1 and $N_{5}$ D5 branes with $N_{1}$ and $N_{5}$
potentially large. This has been in the spirit of time-dependent perturbation
theory of a weakly interacting system: individual interactions are
treated separately and the cumulative effect of many independent interactions
is put together at the
end. We can imagine how the thermal radiation produced by many independent local
quenches can ultimately lead to a thermalized system, but a full, careful treatment remains to be done.
	
We would like to compare the entropies produced from such a process to those of the
system in equilibrium at some finite temperature, not
just the high-temperature limit we considered above.
Computing the finite-temperature entropies would involve
computing two-point functions of twist operators in a domain with both space and time
periodically identified, i.e., a torus. This is somewhat challenging
as the answer would depend on the full operator content
\cite{difrancesco_conformal_1999} and one cannot uniformize to the
plane. It has been carried out for a single fermion
\cite{azeyanagi_near_2008} and we hope to address this for the
D1D5 system in future work. It should also be possible to learn more information
about the state after the quench by judiciously studying the whole set 
of R\'enyi entropies that can be obtained from \eqref{eq:fulldefcorr} and \eqref{eq:wickn} 
(or \eqref{eq:Snbare} for the bare twists),
rather than just $n = 1$ or $2$ as we did here, since they collectively contain
significantly more information. Our results (particularly for the bare,
nonsupersymmetric twist operators)
 may be relevant to and possibly subject to verification by the local 
quenches studied in condensed matter systems, e.g., \cite{2011JSMTE..08..019S, 2007JSMTE, igloi_entanglement_2009, hsu_quantum_2009}. 
However, there are a number of issues in making such a comparison, 
since our CFT and local quench are both highly specific, and 
we have not seriously attempted to do so. 
	
Finally a few words on the bulk description of the process we studied.
The D1-branes are wrapped in the D5-branes, which are localized in the
transverse asymptotically 4+1 dimensional Minkowski space. Initially
the branes are in a stationary state corresponding to the ground state
of the D1D5 orbifold CFT. We then imagine a sudden interaction with an
external field in the transverse space, which weakly deforms the D1D5
system by the exactly marginal twist operator \eqref{eq:def-op} and
supplies the necessary energy for thermalization to begin. This is
admittedly just a sketch that we hope to improve.

\begin{acknowledgments} 
  We thank Tom Faulkner, Joe Polchinski, and Mark Srednicki for
  helpful discussions. CA thanks David Berenstein, Idse Heemskerk,
  Gary Horowitz, SungBin Lee, Don Marolf, and Emil Martinec and SA thanks Borun Chowdhury, Suresh
  Govindarajan, Kalyana Rama, Balachandran Sathiapalan, Masaki
  Shigemori, and Nemani Suryanarayana for conversations and
  suggestions related to this project.  CA is supported in part by the
  DOE under Grant No. DE-FG02-91ER40618.  This research was supported in
  part by the National Science Foundation under Grant No. NSF
  PHY05-51164. SA is grateful for the hospitality and support of the
  KITP, where much of this research was performed.
\end{acknowledgments}


\bibliography{D1D5Therm}
\bibliographystyle{apsrev4-1long}

\end{document}